\documentclass[prc,preprint,showpacs,floatfix,nofootinbib,tightenlines]{revtex4}

\usepackage{graphicx}  %
\usepackage{bm}  %
\usepackage{amsmath,amssymb}

\newcommand{\beqn}{\begin{equation}}
\newcommand{\eeqn}{\end{equation}}
\newcommand{\bea}{\begin{eqnarray}}
\newcommand{\eea}{\end{eqnarray}}
\newcommand{\la}{\langle}
\newcommand{\ra}{\rangle}

\newcommand{\Nmax}{N_{\rm max}}

\newcommand{\flow}{s}
\newcommand{\Trel}{T_{\rm rel}}

\newcommand{\noscr}{|n_1n_2\ra}

\newcommand{\mybar}[1]{\overline{{#1}}}

\newcommand{\vbt}{\overline{V}^{(2)}_s}
\newcommand{\vbtr}{\overline{V}^{(3)}_s}
\newcommand{\vlowk}{V_{{\rm low}\,k}}

\begin{document}


\title{Similarity Renormalization Group Evolution \\ 
of Many-Body Forces in a One-Dimensional Model}

\author{E.D.\ Jurgenson}\email{jurgenson.1@osu.edu}
\author{R.J.\ Furnstahl}\email{furnstahl.1@osu.edu}

\affiliation{Department of Physics,
         The Ohio State University, Columbus, OH\ 43210}

\date{\today}

\begin{abstract}
A one-dimensional system of bosons with short-range repulsion
and  mid-range attraction is used as a laboratory to explore the 
evolution of many-body forces by the Similarity Renormalization
Group (SRG). The free-space SRG is implemented for few-body systems in a
symmetrized harmonic oscillator basis using a recursive
construction analogous to no-core shell model implementations.
This approach, which can be directly generalized to
three-dimensional  nuclei, is fully unitary up to induced
$A$-body forces when applied with an $A$-particle basis (e.g.,
$A$-body bound-state energies are exactly preserved). The
oscillator matrix elements for a given $A$ can then be used in
larger systems. Errors from omitted induced many-body forces show
a hierarchy of decreasing contribution to binding energies.  An
analysis of individual contributions to the growth of many-body
forces demonstrates such a hierarchy and provides an
understanding of its origins.
\end{abstract}
\smallskip

\maketitle


\section{Introduction}

A major goal of nuclear structure theory is to make quantitative
calculations of  nuclear observables starting from  microscopic internucleon
forces.  The Similarity Renormalization Group 
(SRG)~\cite{Glazek:1993rc,Wegner:1994,Kehrein:2006},   through a
continuous series of unitary transformations that soften
initial interactions, can dramatically reduce the computational
requirements of low-energy many-body 
calculations~\cite{Bogner:2006srg,Bogner:2007srg,Bogner:2007rx}.
At the same time, the SRG induces many-body forces as it evolves 
the Hamiltonian. For the SRG to be a useful tool, we must develop
methods for calculating these many-body interactions  and
establish the conditions under which an initial hierarchy of
many-body forces is maintained. In this paper, we study
one-dimensional systems of bosons  as a proof-of-principle of a
practical method to evolve and evaluate such forces,    and
establish a road map for full three-dimensional calculations. 

The SRG can be implemented as a flow equation for the evolving
Hamiltonian $H_\flow$,
\beqn
   \frac{dH_\flow}{d\flow} = 
    [ \eta_s, H_\flow ]
   = [ [G_s, H_\flow], H_\flow] 
    \,.
   \label{eq:commutator}
\eeqn 
Here $s$ is a flow parameter and the flow operator $G_s$
specifies the type of  SRG~\cite{Bogner:2006srg,Anderson:2008mu}.
Most previous applications to nuclear structure have been in a
momentum basis, where decoupling between low-energy and
high-energy matrix elements is naturally  achieved by choosing a
momentum-diagonal flow operator such as the kinetic energy
$\Trel$. However we can evaluate Eq.~(\ref{eq:commutator})  with
$G_s = \Trel$ in any convenient basis.  

In Ref.~\cite{Bogner:2007qb}, a diagrammatic approach to the SRG
equation was introduced, which organized the independent
evolution of two- and three-body (and higher-body) potentials.
This formalism is necessary in a momentum basis to avoid
``dangerous'' delta functions from spectator particles.   An
alternative is to work in a discrete basis for which
Eq.~(\ref{eq:commutator}) can be applied directly in each
$n$-body sector without a separate evolution for few-body forces;
that is, the Hamiltonian is evolved as a whole. We adopt this
approach in the present work, using harmonic oscillator wave
functions as our basis and mimicking the formalism used in the
no-core shell model (NCSM) \cite{NCSM1a,NCSM1b,NCSM1c} to create
properly symmetrized (for bosons) matrix elements in relative
(Jacobi) coordinates. The restriction to one dimension makes the
construction particularly straightforward and requires only
moderate matrix sizes. We use bosons in this paper for easy
comparison with existing model analysis. The boson ground states
coincide with fermion ground states when the flavor degeneracy is
greater than the number of particles, because the overall
anti-symmetrization is realized by the flavor wavefunction. 

We use simple flavor-independent potentials that imitate the
short-range repulsion and mid-range attraction characteristic of
realistic local nuclear potentials. Previous studies of the SRG
imply that properties of the transformations are primarily due to
the matrix structure ($G_s$, $H_s$, choice of basis, etc), so we
expect to be able to directly carry over at least some of our
observations to three dimensions. Because the NCSM formalism is
already developed, the generalization to three-dimensional
fermionic calculations with spin-isospin degrees of freedom and
using realistic nuclear interactions should be algebraically
straightforward (although far more computationally intensive).

The plan of the paper is as follows. In
Section~\ref{sec:formalism}, we develop the one-dimensional
version of the oscillator basis techniques. These include
building symmetrizer operators to construct symmetric basis
states of bosons and fermions, embedding potential and kinetic
energy operators in a given oscillator basis, and developing a
scheme to organize the oscillator basis states in a universal and
scalable manner. In Section~\ref{sec:results}, results are given
for {two-} through five-particle systems. The
two-particle calculations establish that the models simulate the
relevant features of previous nuclear 
calculations~\cite{Bogner:2006srg,Bogner:2007srg,Bogner:2007rx,Jurgenson:2007td}.  
Then three- and higher-particle calculations are used to
explore the running of the few-body potentials.
Section~\ref{sec:conclusion} summarizes our observations and
conclusions.


\section{Formalism}
  \label{sec:formalism}

In this section, we adapt to one dimension the recursive
symmetrization formalism developed by Navratil and collaborators
for use in NCSM calculations with a translationally invariant
harmonic oscillator basis~\cite{NCSM1a,NCSM1b,NCSM1c}. While the
three-dimensional formalism is well documented, the
one-dimensional analog is not, so we provide a self-contained
treatment here.

\subsection{Jacobi Coordinates}

The initial (i.e., unevolved) one-dimensional  Hamiltonian for
$A$ bosons of equal mass $m$ with a local two-body potential has
the first-quantized form (in units with $\hbar=1$)
\beqn
  H = \frac{1}{2m}\sum_{i=1}^Ak_i^2 + \sum_{i<j=1}^AV(x_i-x_j) \;,
  \label{eq:hamiltonian}
\eeqn
where the $x_i$ are single-particle coordinates and the $k_i$ are
single-particle momenta.
To connect to the nuclear problem of interest that uses
potentials given in a momentum basis (e.g., chiral effective field
theory potentials), we calculate matrix elements using harmonic
oscillator basis states in Jacobi momentum coordinates. This
representation also provides a clean visual interpretation of the
SRG evolution of potentials.

With equal-mass particles, a convenient
set of relative momentum Jacobi coordinates is defined by
(for $j = 1 $ to $A-1$)
\beqn
 p_j = \sqrt{\frac{j}{j+1}}\left(\frac{1}{j}\sum_{i=1}^j k_i -k_{j+1}\right) 
 \;,
 \label{eq:Jacobi_coords}
\eeqn
where the $k_i$ are the single-particle momenta of the $A$ particles.
Sometimes for convenience we will use $p\equiv p_1$ and $q\equiv p_2$
when restricted to $A \leq 3$. We define the Fourier transform
$V(x_1-x_2)$ to ${V}(p_1,p_1')$ as
\beqn
  {V}(p_1,p_1') 
    = \int V(\sqrt{2}\ell_1)e^{-i(p_1-p_1')\ell_1} d\ell_1 \;,
  \label{eq:fourier_transform}
\eeqn
where $\ell_1 = (x_1-x_2)/\sqrt{2}$ is the coordinate conjugate
to the Jacobi momentum $p_1$. We introduce a set of harmonic
oscillator states $| n_j \ra$ corresponding to each of the
coordinates of Eq.~(\ref{eq:Jacobi_coords}), so a general product
basis state has the form
\beqn
 \prod_{j=1}^{A-1} \la p_j|n_j\ra \;,
 \label{eq:ho_basis}
\eeqn
with $n_j = 0,1,2,\ldots,\Nmax$ for each $j$. In the next section
we discuss how to build linear combinations of these states that
have the appropriate symmetry.

\subsection{Symmetrization}
\label{sec:symmetrization}

We carry out the SRG evolution for each $A$-particle subsystem in
a complete basis of properly symmetrized states, which will be
linear combinations of the basis states  of
Eq.~(\ref{eq:ho_basis}).   The symmetrization procedure is
adapted from the procedure developed for NCSM
calculations~\cite{NCSM1a,NCSM1b,NCSM1c}. This entails
symmetrizing first the two-particle system and then using a
recursive procedure to go from the $(A-1)$-particle  basis to an
$A$-particle basis. At each stage we keep only symmetric states,
identified as eigenstates of the symmetrizer with eigenvalue
unity.

The two-particle system is specified by the oscillator number
$n_1$.  The symmetrizer is $(1+P_{12})/2$, where $P_{ij}$ is the
exchange operator between particles $i$ and $j$.  Because $P_{12}
|n_1\ra = (-1)^{n_1} |n_1 \ra$, the symmetrizer in the
two-particle case has eigenvalue one acting on states with $n_1$
even and zero when acting on states with $n_1$ odd. Thus the
symmetric basis states have $n_1$ even and we simply omit the
odd states. Following conventions from Ref.~\cite{NCSM1a}, we
label these eigenstates as $|N_2i_2\ra$, where $N_2$ is 
the total oscillator number of the symmetric state and $i_2$ is
an arbitrary label which 
distinguishes states degenerate in $N_2$. In the
two-particle case the notation is trivial, with $N_2 = n_1$ even 
and $i_2 = 1$. We write eigenstate projection coefficients as 
$\la N_2 i_2 \| n_1\ra = \delta_{N_2,n_1}(1+(-1)^{n_1})/2$. These are referred to in
the literature as the ``coefficients of fractional parentage".

A three-particle basis is specified by product states of  the
two-body symmetric  eigenstates, $|N_2i_2\ra$, and
single-particle states with the oscillator number corresponding 
to the third particle, $|n_2\ra$. The symmetrizer for this system
is governed  by the permutation group, $S_3$, which can be
defined by just two of  its generators. Here we choose the
permutation operators $P_{12}$ and $P_{23}$. The symmetrization
operator can be written as
\beqn
  S = \frac{1}{6}(1 + P_{12} + P_{23} + P_{12}P_{23} 
          + P_{23}P_{12} + P_{12}P_{23}P_{12}) \; .
\label{eq:symmetrizer_3N}
\eeqn
We build this symmetrizer in the basis 
$|N_2i_2;n_2\ra \equiv | N_2 i_2 \ra | n_2 \ra$ where the 
states $|N_2i_2\ra$ are already eigenstates of $P_{12}$ with
eigenvalue one, so
Eq.~(\ref{eq:symmetrizer_3N}) reduces to $S = (1+2P_{23})/3$.

In this basis,
the matrix elements of $P_{23}$ can be expressed as 
\beqn
  \la N_2'i_2';n_2'|P_{23}|N_2i_2;n_2\ra 
    = \delta_{N',N}\la n_1'n_2'|n_1n_2\ra_3 \;,
  \label{eq:trans_bracket_3N}
\eeqn
where $N \equiv N_2+n_2 = N_2'+n_2' = N'$ and   $\la
n_1'n_2'|n_1n_2\ra_3$ is the one-dimensional harmonic oscillator
transformation bracket for particles with mass ratio 3.  We
construct these transformation brackets and generalize to mass
ratio $d$ in Appendix~\ref{app:trans_brack}.  By diagonalizing
this symmetrizer we identify the symmetric eigenstates of the
system as the ones with eigenvalue unity.  We keep only those
states and discard the others.  This set of eigenvectors gives us
the coefficients of fractional parentage,  $\la N_2 i_2; n_2 \|
N_3 i_3 \ra$,  of the three-boson symmetric eigenstates, $| N_3
i_3 \ra$, in terms of the original partially symmetrized
three-particle space, $| N_2 i_2; n_2 \ra $.  Note that $i_3$ is
not trivial like $i_2$, because in the three-body system there
are eigenstates degenerate  in the total oscillator number,
$N_2+n_2$.  The label $i_3$ keeps track of those degeneracies. We
find in the one-dimensional system of bosons that the fraction of
symmetric basis states for $A=3$ is about one-fifth. For $A=4$
the reduction in number of states is above 90\%.

To construct the basis states for higher $A$, we generalize this
procedure. To go from $A-1$ to $A$ we need only to symmetrize
between the last two particles, so we construct the symmetrizer
\beqn
  S_A = \frac{1}{A}\bigl(1+(A-1)P_{(A-1)A} \bigr)
  \label{eq:symmetrizer_AN}
\eeqn
in the space of $(A-1)$-particle symmetric eigenstates and the
additional Jacobi state, $n_{A-1}$. We label the basis of
this space as $|N_{A-1} i_{A-1}; n_{A-1}\ra$.  The matrix element
of the exchange operator in this space is
\bea
 && \la N_{A-1}' i'_{A-1}; n_{A-1}'| P_{(A-1)A} | N_{A-1} i_{A-1}; n_{A-1}\ra 
    \nonumber \\
    && \qquad\qquad =   
   \delta_{{N'_{A-1}+n'_{A-1}},{N_{A-1}+n_{A-1}}}\sum
   \la N_{A-1}' i'_{A-1} \| N_{A-2} i_{A-2}; n_{A-2}'\ra 
    \nonumber \\
    &&  \qquad\qquad\quad \null\times
    \la N_{A-2} i_{A-2}; n_{A-2} \| N_{A-1} i_{A-1} \ra
   \la n_{A-2}' n_{A-1}' | n_{A-2} n_{A-1}\ra_{A(A-2)} 
   \;,
  \label{eq:symmetrizer_A}
\eea
where the sum is over $N_{A-2}$, $i_{A-2}$, $n_{A-2}$ and
$n_{A-2}'$. The only significant difference from the
three-particle case is that we must sum over the components of
the $A-1$ subcluster symmetric states to get all the
contributions to the exchange of the last two bosons, $n_{A-2}$
and $n_{A-1}$. The parameter $d=A(A-2)$ can be derived by taking
the last two Jacobi coordinates $p_{A-2}$ and $p_{A-1}$, as
defined in Eq.~(\ref{eq:Jacobi_coords}), and finding the
transformation that exchanges particles labeled by $k_{A-1}$ and
$k_A$. This procedure is shown in Appendix~\ref{app:trans_brack}

For fermions, we need a complete basis of fully anti-symmetrized
states. If we consider the one-flavor case, the procedure for our
one-dimensional model is a trivial modification of
Eqs.(\ref{eq:symmetrizer_3N}) and (\ref{eq:symmetrizer_AN}),
namely all odd permutations come with a minus sign.
Thus, for $A=3$ the anti-symmetrizer can be written
\beqn
A = \frac{1}{3}(1-2P_{23}) \;,
\eeqn
where $P_{23}$ acts on the flavor space as well. If there are
more flavors than particles and the interaction is flavor
independent, the spatial wavefunction for the ground state will
be symmetric and correspond to our boson ground state
wavefunctions. For realistic three-dimensional nuclei, the
required construction of an anti-symmetric Jacobi basis with full
angular momentum coupling has been worked out for the NCSM by
Navratil et al.~\cite{NCSM1a,NCSM1b,NCSM1c}.

\subsection{Hamiltonian Matrix Elements}
\label{sec:hamiltonian}

To obtain the Hamiltonian in the symmetric eigenbasis of the
general $A$-particle system, we employ a recursive embedding
procedure that utilizes the partially symmetric bases developed
for the symmetrization operator. First we treat the kinetic
energy and then the potential.

The relative kinetic energy in the three-particle system is the total 
minus the center-of-mass kinetic energies:
\bea
  T_{\rm rel} &\equiv& T_{\rm tot} - T_{\rm cm} \nonumber \\ 
  &=& \frac{k_1^2}{2m} + \frac{k_2^2}{2m} + \frac{k_3^2}{2m} 
  - \frac{(k_1+k_2+k_3)^2}{2(3m)} \nonumber \\ 
  &=& \frac{p_1^2 + p_2^2}{2m} \;,
  \label{eq:Trel}
\eea
where the $p_i$'s are defined in  Eq.~(\ref{eq:Jacobi_coords}).
Momentum basis states are organized by  increasing kinetic
energy.  We can project $T_{\rm rel}$ directly onto the
three-particle oscillator basis by using the ladder operator
definitions of the Jacobi momenta. The  projection of $T_{\rm
rel}$ into the $\noscr$ basis is
\bea
\la n_1' n_2' | T_{\rm rel} |n_1n_2\ra
  &=&  \la n_1' n_2'|\frac{p_1^2 + p_2^2}{2m} |n_1n_2\ra
    \nonumber \\ 
  &=&  \frac{1}{2m} \frac{-m\omega}{2} 
  \la n_1' n_2'|(\eta_1^\dagger - \eta_1)^2 + (\eta_2^\dagger - \eta_2)^2 |n_1n_2\ra 
 \;,
\eea
where the $\eta_1$ and $\eta_2$ operators act on the $n_1$ and $n_2$ spaces, 
respectively.  Continuing, we get
\bea
\la n_1' n_2' | T_{\rm rel} |n_1n_2\ra  &=& \frac{1}{2m} \frac{-m\omega}{2} 
    [\la n'_1 | (\eta_1^\dagger - \eta_1)^2 | n_1 \ra \delta_{n_2,n'_2} 
    + \la n'_2 | (\eta_2^\dagger - \eta_2)^2 | n_2 \ra \delta_{n_1,n'_1}] 
   \nonumber \\ 
&=& \frac{-\omega}{4} 
          \Bigl[ \bigl( \sqrt{(n_1+1)(n_1+2)}\,\delta_{n'_1,n_1+2} 
   + \sqrt{n_1(n_1-1)}\,\delta_{n'_1,n_1-2} 
   \nonumber \\ 
   & & \qquad \null 
      - (2n_1+1)\delta_{n'_1,n_1}
   \bigr)\, \delta_{n'_2,n_2} 
   \nonumber \\ 
 && \qquad \null + \bigl( \sqrt{(n_2+1)(n_2+2)}\,\delta_{n'_2,n_2+2} 
   + \sqrt{n_2(n_2-1)}\,\delta_{n'_2,n_2-2}
   \nonumber \\ 
   & & \qquad \null 
    - (2n_2+1)\delta_{n'_2,n_2} 
    \bigr)\, \delta_{n'_1,n_1}\Bigr] \;.
  \label{eq:T_exact}
\eea
As noted, we keep only the $n_1$-even states using the projector
$\la N_2 i_2 \| n_1\ra$, and we can symmetrize the full
three-particle system with the symmetric eigenstates, $|N_3i_3\ra$
 whose components are given by $\la N_3i_3\|N_2i_2;n_2\ra$. 

To derive the $A$-body kinetic energy in the symmetrized basis,
$(T_A)_{\rm sym}$, we use a recursive procedure on the
$(A-1)$-body result to find the $A$-particle space operator
matrix elements: 
\bea
  (T_{A})_{\rm sym} 
         &=& \la N_A' i'_A|T_A|N_A i_A\ra 
	 \equiv \la N_A' i'_A| \sum_{i=1}^{A-1} p_i^2/2m |N_A i_A\ra \nonumber \\
         &=& \la N_A'\|N_{A-1}'n_{A-1}'\ra
             \la N_{A-1}'n_{A-1}'|T_A|N_{A-1}n_{A-1}\ra
             \la N_{A-1}n_{A-1}\|N_A\ra 
             \nonumber \\
         &=& \la N_A'\|N_{A-1}'n_{A-1}'\ra
             \bigl[
              \la N_{A-1}'n_{A-1}'|(T_{A-1})_{\rm sym}+p_{A-1}^2/2m|N_{A-1}n_{A-1}\ra
             \bigr]
             \la N_{A-1}n_{A-1}\|N_A\ra 
             \nonumber \\
         &=& \la N_A'\|N_{A-1}'n_{A-1}'\ra
             \bigl[
              \la N_{A-1}'|(T_{A-1})_{\rm sym}|N_{A-1}\ra \delta_{n_{A-1}',n_{A-1}} 
             \nonumber \\
         && \hspace{1cm}  \null + \delta_{N_{A-1}',N_{A-1}}
               \la n_{A-1}'|p_{A-1}^2/2m|n_{A-1}\ra
              \bigr]
              \la N_{A-1}n_{A-1}\|N_A\ra 
             \nonumber \\
         &=& \la N_A'\|N_{A-1}'n_{A-1}'\ra
             \bigl[(T_{A-1})_{\rm sym}\delta_{n_{A-1}',n_{A-1}}
             - \frac{\omega}{4}\delta_{N_{A-1}',N_{A-1}}
             \bigl( \delta_{n_{A-1}',n_{A-1}}(2n_{A-1}+1) 
             \nonumber \\
         && \hspace{1cm}  
              - \delta_{n_{A-1}'+2,n_{A-1}}\sqrt{n_{A-1}^2 - n_{A-1}} 
              - \delta_{n_{A-1}'-2,n_{A-1}}\sqrt{(n_{A-1}+1)(n_{A-1}+2)} 
              \,\bigr)\bigr]
              \nonumber \\ 
         &&  \null\times \la N_{A-1}n_{A-1}\|N_A\ra  \; ,
	     \label{eq:T_AS}
\eea
where we have suppressed the $i_A$'s and $i_{A-1}$'s for
simplicity after the first line.  Intermediate summations over
$N_{A-1}$, $n_{A-1}$, $i_{A-1}$, $N_{A-1}'$, $n_{A-1}'$, and
$i_{A-1}'$ are implicit.

In the same manner as the kinetic energy, we can recursively
embed the potential in the $A$-particle space, starting with the
two-body interaction between the first two particles. Because we
are working in fully symmetrized few-body spaces we do not need
to consider all pair-wise interactions, but only one such pair
and scale by the number of interactions. For instance, in the
three-particle system the full two-body interaction is $V^{(2)} =
V_{12} + V_{23} + V_{13} = 3 V_{12}$. In a general $A$-particle
space, this becomes $V^{(2)} = {A \choose 2} V_{12}$. The matrix
element of a two-body potential, $V_{12}$, in the relative
coordinate harmonic oscillator basis, $|n_1\ra$, is $\la n_1'|
V_{12} | n_1\ra = \int \la n_1'| p_1' \ra \la p'| V_{12} | p \ra
\la p |n_1\ra dp\, dp'$ where the matrix elements of $\la
p'|V_{12}|p\ra$ are given by  Eq.~\eqref{eq:fourier_transform}. 

Once in the oscillator basis, embedding in a larger particle
space is a straightforward process. Starting with the two-body
interaction, $V_{12}$, the two-body oscillator symmetric states
are isolated using the projector $\la N_2 i_2|n_1 \ra$ which picks
out just the $n_1$-even states. Embedding this interaction in the
three-particle space involves adding a new Jacobi coordinate,
$|n_2\ra$, to the existing system. With respect to the two-body
interaction, $V_{12}$, this additional coordinate is associated
with a delta function, $\delta_{n_2,n_2'}$. Finally we obtain the
symmetric three-particle states by using  the projector,  $\la
N_3i_3||N_2i_2;n_2\ra$. Multiplying by ${3 \choose 2} = 3$ gives
us the full strength of the two-body interaction.

In general we can write this procedure as an
expansion of the final $A$-particle symmetric space matrix
elements of $V_{12}$:
\bea
  (V^{(2)}_{A})_{\rm sym} &=& \la N_A' i'_A|V^{(2)}_A|N_A i_A\ra 
  	\equiv  {A \choose 2} \la N_A' i'_A|V_{12}|N_A i_A\ra \nonumber \\
         &=& \la N_A'\|N_{A-1}'n_{A-1}'\ra
             \la N_{A-1}'n_{A-1}'|V_A|N_{A-1}n_{A-1}\ra
             \la N_{A-1}n_{A-1}\|N_A\ra 
             \nonumber \\
         &=& \la N_A'\|N_{A-1}'n_{A-1}'\ra \nonumber \\
	 && \null \times \la N_{A-1}'n_{A-1}'|(V_{A-1})_{\rm sym}
	     \delta_{n_{A-1}',n_{A-1}}|N_{A-1}n_{A-1}\ra
             \la N_{A-1}n_{A-1}\|N_A\ra  
\label{eq:v_osc_A}
\eea
where again we have dropped the $i_A$'s after the first line for
simplicity and intermediate sums are implicit. We remind the
reader that $n_{A-1}$ can only take values from $0$ to $N -
N_{A-1}$, where $N$ is the total oscillator quantum number used
to organize the states. We start with the two-particle space and
work our way up to the $A$-body space, embedding the interactions
successively in each sector using Eq.~(\ref{eq:v_osc_A}). 
When symmetrizing $V_A$ we must embed the symmetrized $V_{A-1}$
with the appropriate combinatoric factor included as explained
above. This factor derives from the fact that we had embedded a
2-body force in the $A-1$ space that is now to be extracted and
embedded in the $A$-particle space. Thus we must remove the old
factor $A-1 \choose{2}$ and multiply by the new $A \choose{2}$
factor,  which has the net effect of  multiplying by $A/(A-2)$. 

Any initial three-body force (discussed below) is embedded in the
same manner as above except that it originates in the
three-particle space. The initial three-body force is a function
of two Jacobi momenta, which we transform directly into the
partially symmetrized three-particle oscillator space and then
use all of the same embedding procedures developed above. Note
that two- and three-body forces must be embedded in higher spaces
with different symmetry factors, ${A \choose 2}$ and ${A \choose
3}$ respectively. 

In previous formulations of this recursive approach~\cite{NCSM1a},
subsequent potential embeddings are achieved by making a change
of coordinates for the last two Jacobi momenta. For systems with
$A>5$, the three-body force requires a similar change of
coordinates for the last three Jacobi momenta. Such a scheme 
is unnecessary here.

\subsection{SRG Evolution}

Once we have constructed a complete symmetrized basis for $A$
particles (specified by the value of $\Nmax$) and evaluated the
Hamiltonian matrix elements in this basis,
applying Eq.~(\ref{eq:commutator}) with $G_s = T_{\rm rel}$ is
immediate. That is, we have coupled, first-order differential
equations  for each matrix element of the Hamiltonian, with the
right side of each equation given by a series of matrix
multiplications. This is efficiently implemented in any computer
language with matrix operations and differential equation
solvers.

Individual matrix elements of the Hamiltonian obey the 
SRG's differential equations:
\bea
\frac{d}{ds}\la N_A'i_A'|(V_A)_s|N_Ai_A\ra &=&
\la N_A'i_A'|\big[\big[T,H_s\big],H_s\big]|N_Ai_A\ra \nonumber \\
&=& \la N_A'i_A'|T H_s H_s|N_Ai_A\ra 
+ \la N_A'i_A'|H_s H_s T|N_Ai_A\ra \nonumber \\
 & & \null - 2\la N_A'i_A'|H_s T H_s|N_Ai_A\ra 
 \;.
 \label{eq:SRGmatrices} 
\eea
We have defined ${dT}/{ds}=0$ so that all of the flow occurs in
the matrix representation of the potential, $(V_A)_s$.  Using the
matrix representations of $T$ and $H_s$ in the $|N_A i_A\ra$
basis, the right side of Eq.\eqref{eq:SRGmatrices} is simply a
series of matrix multiplications. The initial condition at $s=0$
is the initial Hamiltonian, $\la N_A'i_A'|T+V_A|N_Ai_A\ra$, which
can have few-body components in $V_A$. We consider here both
two-body-only and two-body plus a three-body component (for $A
\geq 3$). In practice we often use the flow variable $\lambda =
1/s^{1/4}$; because there is no explicit $s$ dependence in
Eq.~(\ref{eq:commutator}), switching variables is trivial. 
 
To carry out the SRG evolution, we use a built-in MATLAB
differential equation solver, such as the MATLAB function  {\tt
ode23}, which is an implementation of a Runge-Kutta differential
equation algorithm.  We studied the running time to evolve the
various SRG schemes (i.e., the choice of $G_s$) by plotting the
time to run versus the evolution parameter, $s$. We find a
straight line as $s$ increases, indicating no stiffness, in every
combination of potential and SRG scheme used to date. 

The SRG induces few-body forces as it evolves an initial
interaction in a few-particle space. To study the contributions
of different few-body forces we must isolate these components of
the full interaction. By definition the two-body force evolution
keeps the $A=2$ binding energy invariant under evolution. We can
isolate the three-body force from the two-body matrix elements by
embedding the evolved two-body-only force in the three-particle
space and subtracting it from the full two-plus-three-body
evolved interaction. In our MATLAB implementation these
procedures take only a few lines of code.


\section{results}
  \label{sec:results}

\subsection{Initial (``Bare") Interactions}

\begin{figure}[tbh!]
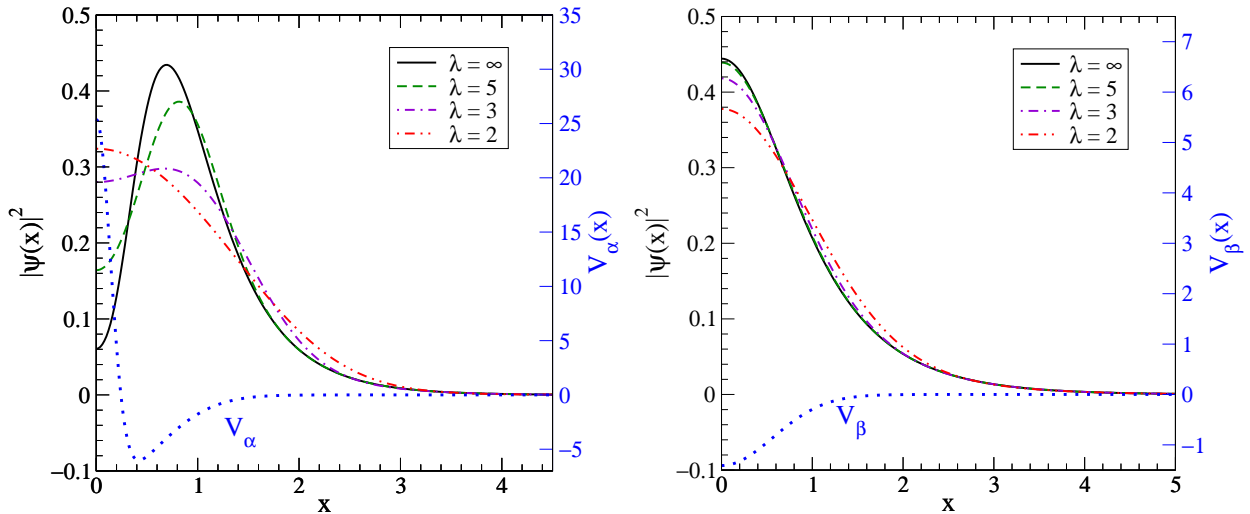

\begin{center}
 \includegraphics*[width=3.2in]{wf_test_Va_C0}
 \hfill
 \includegraphics*[width=3.2in]{wf_test_Vf_C0}
\end{center}
\vspace*{-0.1in}
\caption{(color online) Potentials (dotted line, with axis on
  right)  and probability distributions (other lines, with axis
  on left) for the lowest two-body bound state as a function of
  $x = |x_1 - x_2|$ at different stages in the SRG evolution
  ($\lambda = 1/s^{1/4}$). The left plot is $V_{\alpha}$ and the
  right plot is  $V_{\beta}$ (see Table~\ref{tab:negelePars}).}
\label{fig:srg_2_body}
\end{figure}

The bulk of our calculations adopt a model from Ref.~\cite{negelePot}
that uses a sum of two gaussians to simulate repulsive
short-range and attractive mid-range nucleon-nucleon two-body
potentials:
\beqn
  V^{(2)}(x) = \frac{V_1}{\sigma_1\sqrt{\pi}} e^{-x^2/\sigma_1^2}
    + \frac{V_2}{\sigma_2\sqrt{\pi}} e^{-x^2/\sigma_2^2} 
\eeqn
or
\beqn
   V^{(2)}(p,p') = \frac{V_1}{2 \pi\sqrt{2}}e^{-(p-p')^2 \sigma_1^2/8} 
          + \frac{V_2}{2 \pi\sqrt{2}}e^{-(p-p')^2 \sigma_2^2/8} \;.
  \label{eq:gaussians}
\eeqn
The parameters used in Ref.~\cite{negelePot} were chosen  so that
the one-dimensional saturation properties correspond to empirical
three-dimensional properties, but we also want to explore a range
of parameters to test what behavior is general and what relies on
specific features.  We start with the parameters listed in
Table~\ref{tab:negelePars}. The potential  $V_{\alpha}$ is from
Ref.~\cite{negelePot} and is plotted in
Fig.~\ref{fig:srg_2_body}. We will fix the range of the
attractive part and vary the relative strength and range of the
repulsive parts and visa versa.  We also vary the purely
attractive potential $V_{\beta}$, which was used in
Ref.~\cite{VanNeck:1996} and is also plotted in
Fig.~\ref{fig:srg_2_body}. The eigenvalue problem for the
relatively small matrices considered here can be solved by any
conventional matrix diagonalization program (MATLAB was used
here).

\begin{table}[tb]
\caption{Parameter sets for the two-body potential of 
Eq.~(\ref{eq:gaussians}). }
\begin{tabular}{c|cccc}
name & $V_1$ & $\sigma_1$ & $V_2$ & $\sigma_2$ \\
\hline
$V_{\alpha}$ & 12. & 0.2 & $-12.$ & 0.8 \\
$V_{\beta}$ & 0. & 0.0 & $-2.0$ & 0.8 \\
\end{tabular}
\label{tab:negelePars}
\end{table}

To test that the symmetrized harmonic oscillator basis was
correctly constructed for $A = 2$, $3$, and $4$ (see below), we
first diagonalized the Hamiltonian using the purely attractive
gaussian  two-body potential $V_{\beta}$.  The normalization is
such that $V_{\beta}(x)$ becomes a delta function with strength
$V_2$ as $\sigma_2 \rightarrow 0$~\cite{VanNeck:1996} (note the
numerical factors from the Fourier transform because of our
normalization of the Jacobi momenta). This limiting case has a
known analytic solution for the (only) bound state of $A$
bosons.  For finite $\sigma_2$, we were able to confirm the
accuracy of the diagonalizations as a function of the basis size
$\Nmax$ by comparison to  coordinate-space stochastic variational
method (SVM) calculations using a published code~\cite{svmcpc}
adapted to one dimension~\cite{EricPC}.

\newcommand{\ph}{\phantom{-}}

\begin{table}[tb]
 \caption{Ground-state energies for two-body potentials from
 Table~\ref{tab:negelePars} with various strengths of the
 initial three-body potential
 Eqs.~\eqref{eq:three_body_force}--\eqref{eq:fLambda} 
  with $\Lambda = 2$ and $n=4$ 
 for $A = 2$, 3, and 4.}
 \begin{tabular}{cc|ccc}
   $V^{(2)}$ &  $c_E$  &  $E_2$  & $E_3$  &  $E_4$  \\
   \hline
$V_{\rm \alpha}$ & $-0.10 $ & $ -0.920 $ & $ -3.223 $ & $ -7.125 $ \\
$V_{\rm \alpha}$ & $-0.05 $ & $ -0.920 $ & $ -2.884 $ & $ -5.832 $ \\
$V_{\rm \alpha}$ & $-0.01 $ & $ -0.920 $ & $ -2.628 $ & $ -4.906 $ \\
$V_{\rm \alpha}$ & $ 0.00 $ & $ -0.920 $ & $ -2.567 $ & $ -4.695 $ \\
$V_{\rm \alpha}$ & $ 0.01 $ & $ -0.920 $ & $ -2.507 $ & $ -4.494 $ \\
$V_{\rm \alpha}$ & $ 0.05 $ & $ -0.920 $ & $ -2.278 $ & $ -3.798 $ \\
$V_{\rm \alpha}$ & $ 0.10 $ & $ -0.920 $ & $ -2.027 $ & $ -3.179 $ \\
$V_{\rm \beta}$  & $-0.10 $ & $ -0.474 $ & $ -3.379 $ & $ -8.412 $ \\
$V_{\rm \beta}$  & $-0.05 $ & $ -0.474 $ & $ -2.283 $ & $ -5.727 $ \\
$V_{\rm \beta}$  & $-0.01 $ & $ -0.474 $ & $ -1.792 $ & $ -4.183 $ \\
$V_{\rm \beta}$  & $ 0.00 $ & $ -0.474 $ & $ -1.708 $ & $ -3.846 $ \\
$V_{\rm \beta}$  & $ 0.01 $ & $ -0.474 $ & $ -1.626 $ & $ -3.517 $ \\
$V_{\rm \beta}$  & $ 0.05 $ & $ -0.474 $ & $ -1.370 $ & $ -2.451 $ \\
$V_{\rm \beta}$  & $ 0.10 $ & $ -0.474 $ & $ -1.240 $ & $ -1.874 $
\end{tabular}   
 \label{tab:gs_energies}
\end{table} 

We also explore the evolution of Hamiltonians with an initial
three-body force. We choose a regulated contact interaction in
the three-particle momentum space,
\beqn
 V^{(3)}(p,q,p',q') =  c_E f_\Lambda(p,q) f_\Lambda(p',q') \; ,
\label{eq:three_body_force}
\eeqn
where $c_E$ is the strength of the interaction and the
form factor $f_\Lambda$ depends on the Jacobi momenta as
\beqn
  f_\Lambda(p,q) \equiv e^{-((p^2+q^2)/\Lambda^2)^n} \;.
  \label{eq:fLambda}
\eeqn
The regulator cutoff $\Lambda$ sets the scale of the fall-off in
momentum and $n$ determines the sharpness of this fall-off. This
form is analogous to the regulated three-body contact
interactions used in chiral effective field
theory~\cite{Epelbaum:2002vt}. We have not explored in detail the
impact of adjusting $\Lambda$ and $n$ but have focused on how the
SRG handles a varying strength $c_E$. All results here are for
$\Lambda = 2$ and $n=4$.

Most of the figures in this paper show calculations with $\Nmax =
28$. With this basis size, the ground-state energies   are
generally converged to one part in $10^{4}$, which is more than
sufficient for our purposes.  As usual, increasing $\Nmax$ leads
to rapidly increasing matrix sizes and computation times;  times
for $A=3$ with $\Nmax = 32$ are a factor of 3 longer than with
$\Nmax = 28$ and with $\Nmax = 40$ the time increases by another
factor of 10. A sampling of ground-state energies are given in
Table~\ref{tab:gs_energies}.  

\subsection{Two-body Results}

We first consider the bound state of two identical bosons using
the potential $V_{\alpha}$. Because the SRG is a series of
unitary transformations, we expect that the binding energy will
not be changed by evolving the two-body interaction in the
two-particle space.  Indeed, we find it to be constant to high
accuracy. The ground-state wave function, however, changes
dramatically, as seen from the probability densities plotted in
Fig.~\ref{fig:srg_2_body}. The initial probability density
exhibits a sizable ``wound'' near the origin that is filled in as
$\lambda$ decreases.  By $\lambda = 2$ there is no signature of a
repulsive core (and the wave function is modified out to larger
$x$). This is the same pattern seen for the S-wave component of
deuteron wave functions starting from three-dimensional
nucleon-nucleon S-wave potentials with strong repulsive cores
such as Argonne $v_{18}$, with the ``uncorrelated'' final wave
function at $\lambda = 2$ roughly comparable to $\lambda =
1.5\,\mbox{fm}^{-1}$ for the deuteron~\cite{Bogner:2007srg}.

\begin{figure}[tbh!]
\begin{center}
 \includegraphics*[height=1.5in]{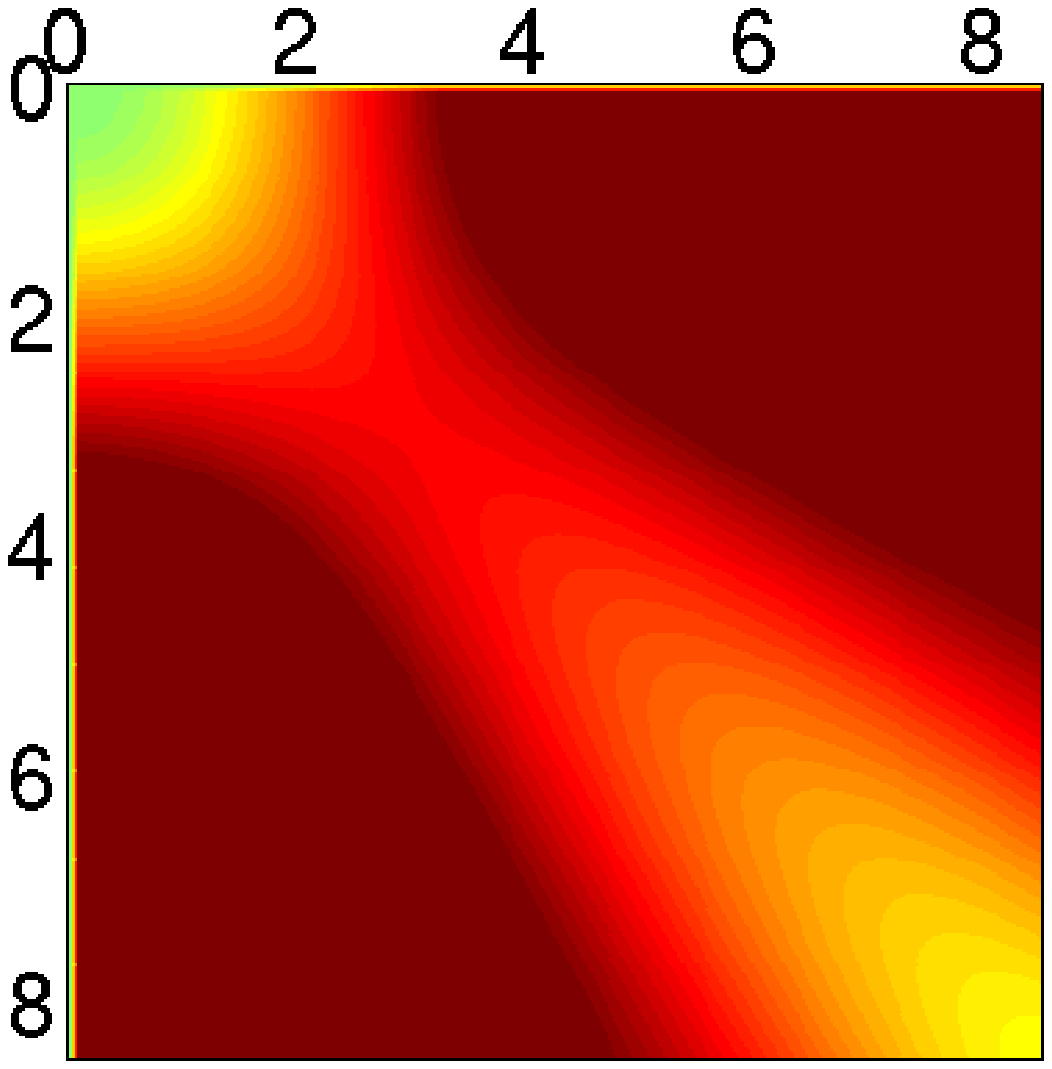}
\includegraphics*[height=1.5in]{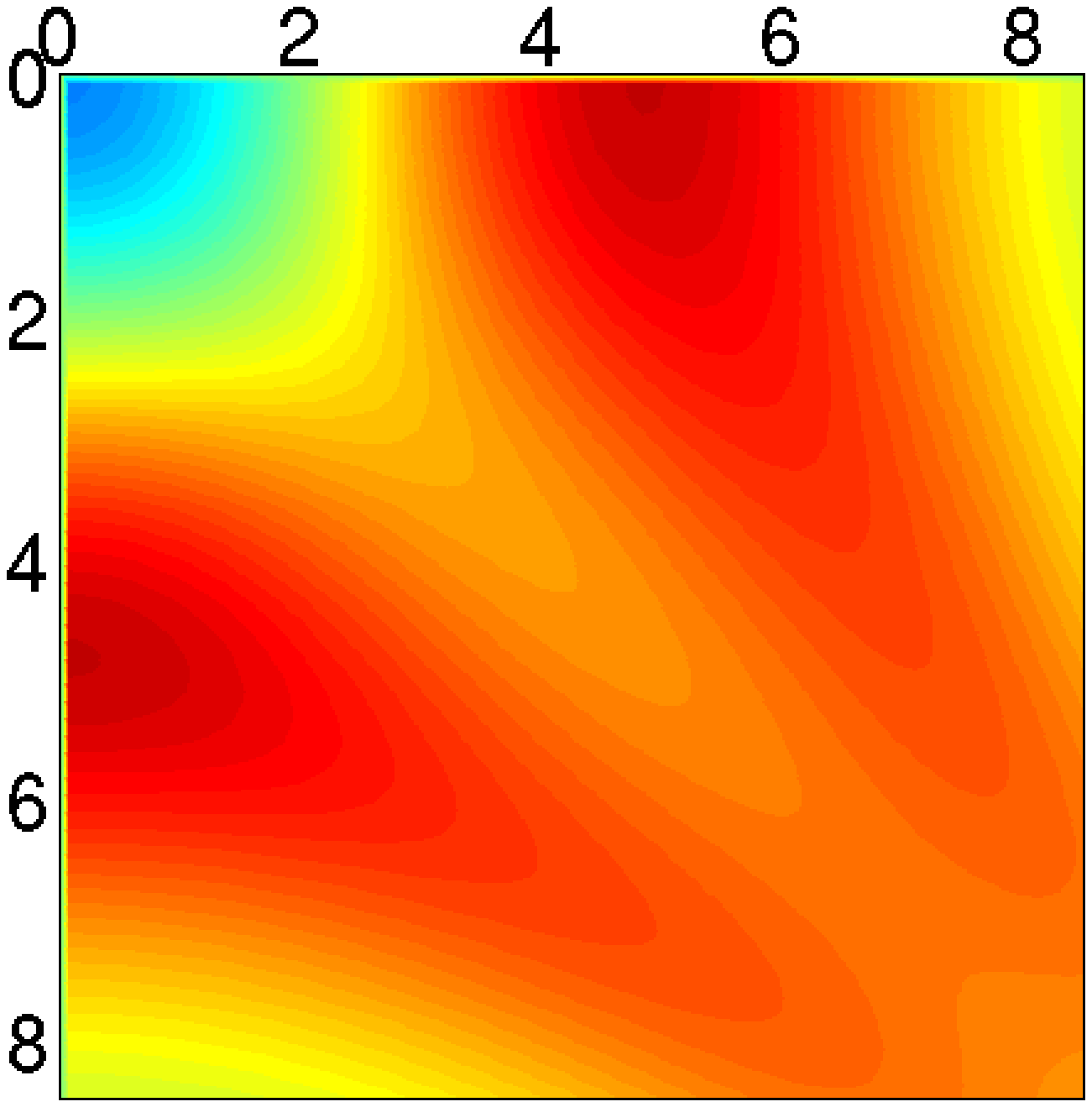}
 \includegraphics*[height=1.5in]{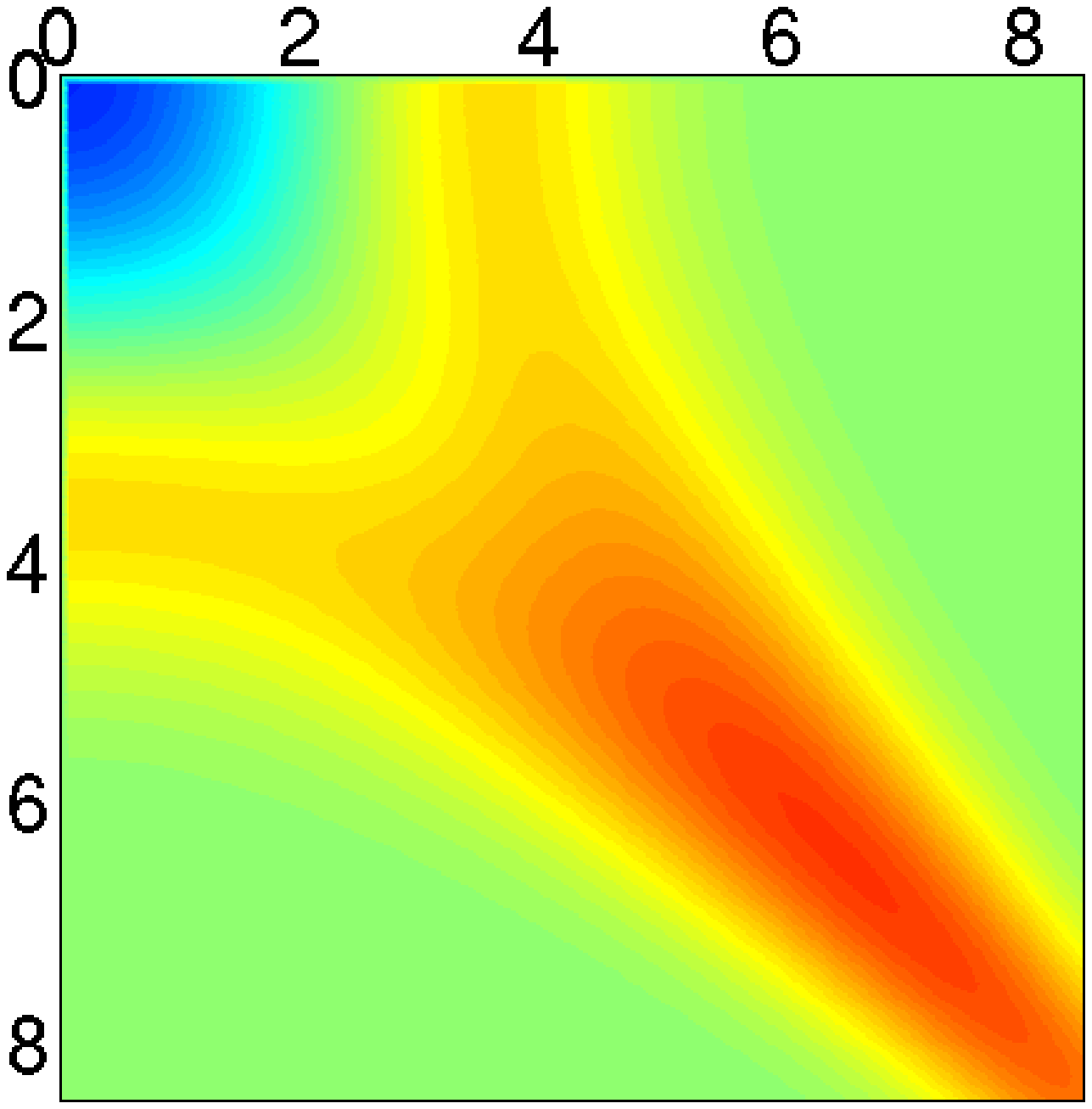}
 \includegraphics*[height=1.5in]{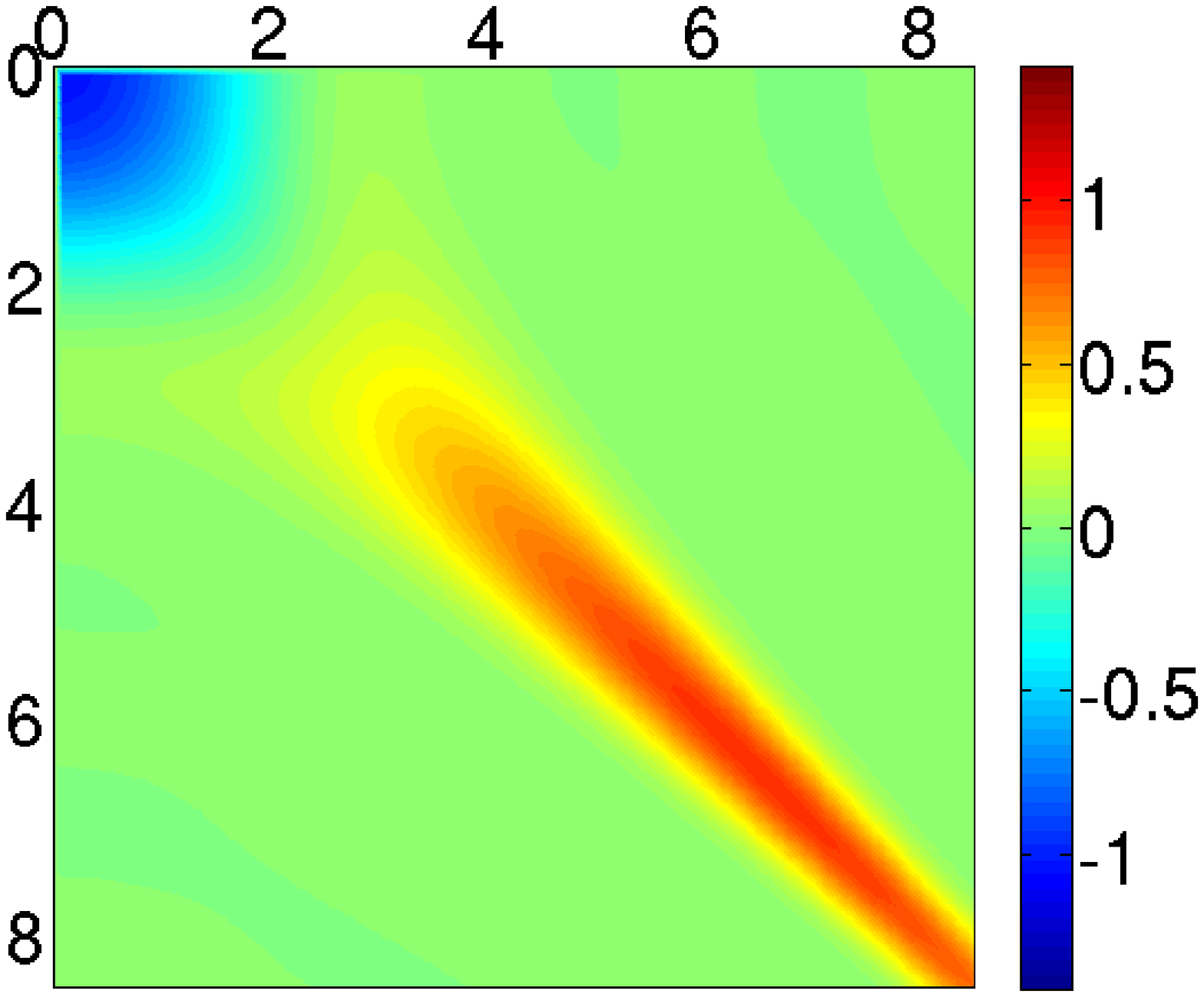}
\end{center}
\vspace*{-0.1in}
\caption{(color online) Even part of the SRG potential  $
[V^{(2)}_{s}(p,p') + V^{(2)}_{s}(p,-p')]$ in dimensionless
units as a function of $p$ and $p'$ for  $\lambda = \infty$, 5,
3, and 2 (where $\lambda = 1/s^{1/4}$).  The initial potential is
$V_{\alpha}$ from Table~\ref{tab:negelePars}.}
\label{fig:srg_2_body_mom}
\end{figure}

The evolution of the potential in the momentum basis, shown as a color
contour plot in Fig.~\ref{fig:srg_2_body_mom}, also demonstrates
this behavior. (The even part of the potential is shown, which is
the analog of the S-wave part.) The initial potential is
dominated by strongly repulsive matrix elements coupling low and
high momenta.  The evolution in $\lambda$ band diagonalizes the
potential to a width in $p^2$ of roughly $\lambda^2$ while a soft
attractive part emerges in the low-momentum region. The pattern
in Fig.~\ref{fig:srg_2_body_mom} reflects increasing non-locality
as $\lambda$ is lowered, which in turn reduces the wound in the
wave function. From the probability density and the momentum
space plots we estimate that evolving to halfway between $\lambda
= 2$ and 3 for $V_{\alpha}$  corresponds roughly to the $\lambda$
scale typically used in nuclear structure calculations (around
$2\,\mbox{fm}^{-1}$).

\subsection{Three-body Results}

To calculate properties of the three-particle system we construct
the Hamiltonian in the basis of symmetric three-particle
eigenstates as described in Sec.~\ref{sec:hamiltonian}. The SRG
evolution of the potential in the three-particle space leaves the
ground state energy invariant if the full Hamiltonian is kept,
because the transformations are unitary. However,  the
Hamiltonian matrix elements in this space  do not follow simply
from the pairwise sum of the two-body potential matrix elements; 
as the SRG evolves a three-body force is induced even if its
initial strength is zero.

\begin{figure}[tbh!]
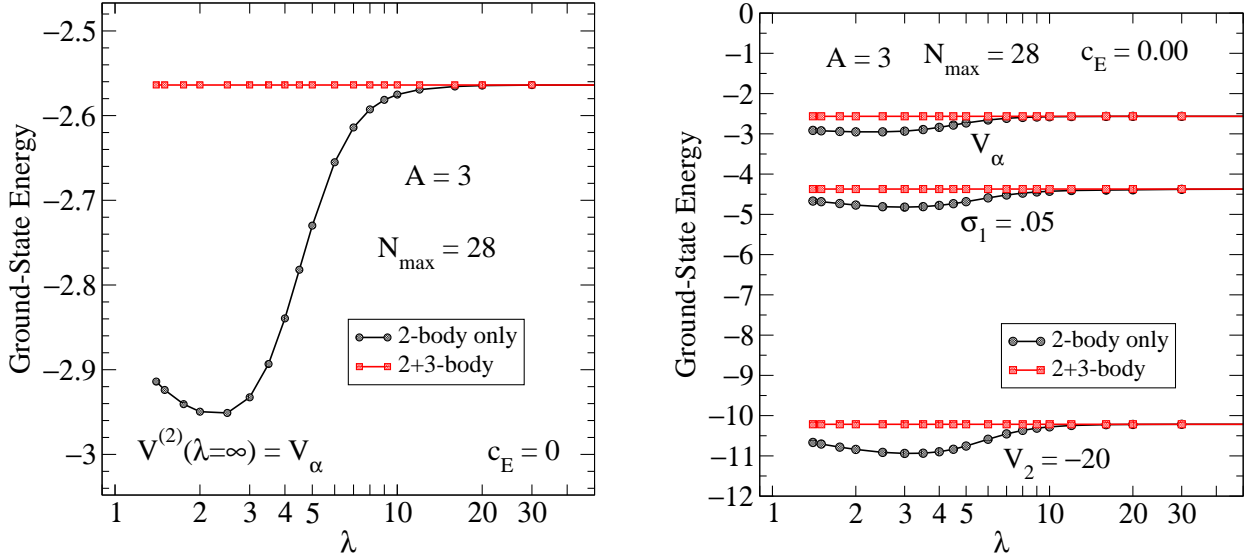

\begin{center}
 \includegraphics*[height=2.9in]{Ebind_3N_Va.eps}
  \hfill
 \includegraphics*[height=2.9in]{Ebind_3N_Vamulti.eps}
\end{center}
\vspace*{-0.1in}
\caption{(color online) The lowest bound-state energy $E_3$ for a
three-particle system as a function of $\lambda$ with the initial
two-body-only potential $V_{\alpha}$.   The (red) curves with
squares include the full evolution of the Hamiltonian while the
(black) curves with circles use the two-body potential evolved
in the two-particle system.  The right frame shows two additional
results from varying $\sigma_1$ and $V_2$ from the values in
Table~\ref{tab:negelePars}.}
\label{fig:srg_3_body}
\end{figure}

\begin{figure}[tb]
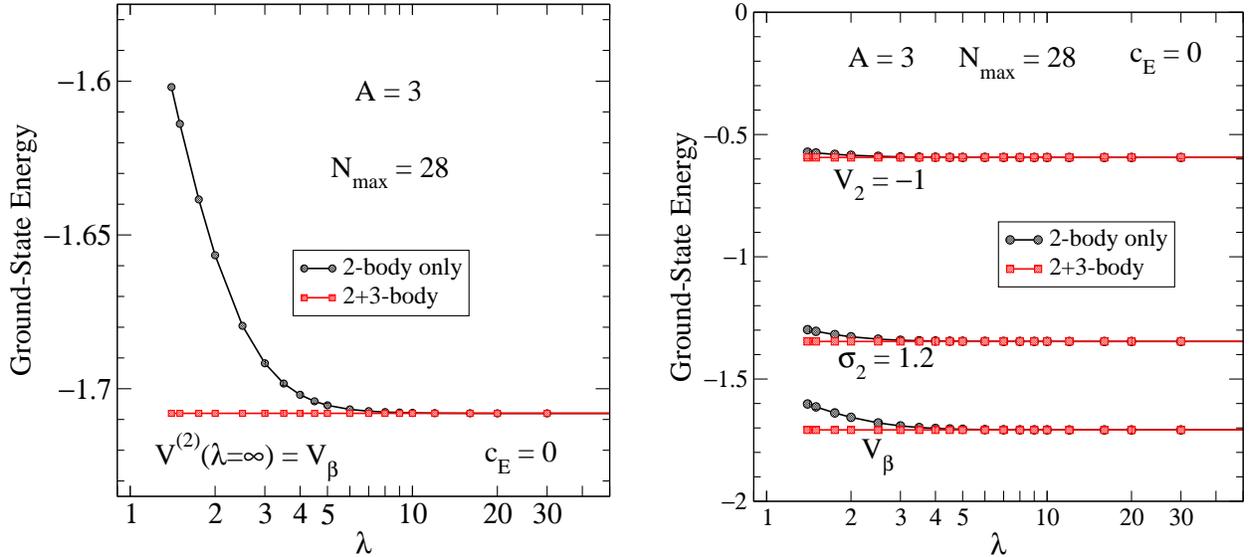

\begin{center}
  \includegraphics*[height=2.9in]{Ebind_3N_Vf.eps}
  \hfill
  \includegraphics*[height=2.9in]{Ebind_3N_Vfmulti.eps}
\end{center}
\vspace*{-0.1in}
\caption{(color online) Same as Fig.~\ref{fig:srg_3_body} but
  with initial potential $V_{\beta}$.  The right frame shows two
  additional results from varying  $\sigma_2$ and $V_2$ from the
  values in Table~\ref{tab:negelePars}.}
\label{fig:3N_other}
\end{figure}

The effect of this full three-particle space SRG evolution is
shown in Fig.~\ref{fig:srg_3_body} for initial two-body potential
$V_{\alpha}$ and with initial $V^{(3)} = 0$ (non-zero values of
$c_E$ are considered in the next section). We plot the
ground-state energy for the three-particle system both with the
initial two-body interaction embedded in the three-particle
symmetric space and then evolved (the red curve with squares) and
also with the two-body interaction evolved in the two-particle
space before embedding in the three-particle space at each
$\lambda$ (the black curve with circles).  We can see that the energy
evaluated with the two-body interaction alone deviates noticeably
as $\lambda$ drops below 10. This variation is the signature that the
two-body transformation is only approximately unitary in the
three-particle sector. The error reaches a peak in $\lambda$
between 2 and 3 and then decreases. The same pattern has been
observed for NN potentials in three
dimensions~\cite{Bogner:2007rx} and remains qualitatively the
same when parameters in the potential are varied (e.g., see the
right plot in Fig.~\ref{fig:srg_3_body}). We made the same
calculation using the purely attractive initial two-body
potential $V_{\beta}$, which is shown in Fig.~\ref{fig:3N_other}.
Here the induced three-body force has the opposite sign and there
is no  maximum, which implies that the qualitative pattern of
evolution is dictated by the interplay between attractive
long-range and repulsive short-range parts of the potential. 
These features are explored further in
Sect.~\ref{subsec:analysis}.

\begin{figure}[tb]
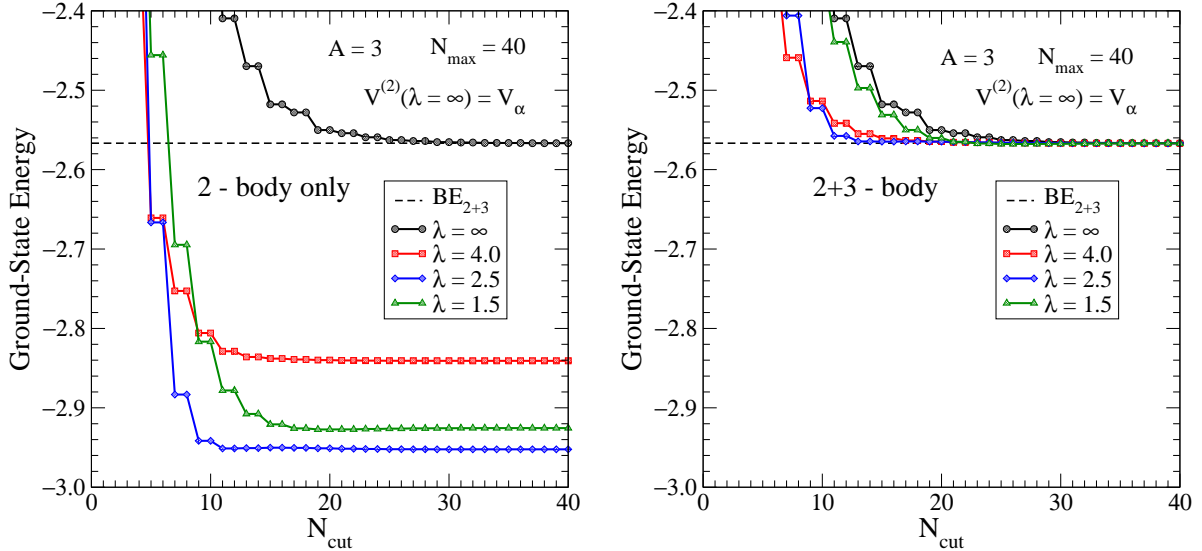

\begin{center}
  \includegraphics*[width=3in]{decoupling_2body.eps}
  \hspace*{.1in}
  \includegraphics*[width=3in]{decoupling_3body.eps}
\end{center}
\vspace*{-0.1in}
\caption{(color online) Decoupling in the three-particle system. 
The intial $V_{\alpha}$ potential is evolved to each $\lambda$
shown in a basis with $\Nmax = 40$.   On the left only the
two-body potential is kept while the full potential is kept on
the right. Matrix elements of the potential are set to zero if
one or both states have $N > \mbox{$N_{\rm cut}$}$ and the resulting
Hamiltonian is diagonalized to obtain the ground-state energies
plotted.}
\label{fig:srg_3_body_decoupling}
\end{figure}

In Fig.~\ref{fig:srg_3_body_decoupling} we test SRG decoupling
\cite{Jurgenson:2007td} within the harmonic oscillator basis.  An
initial two-body potential is evolved in a large $A=3$ space both
with and without the induced  three-body interaction. Then the
Hamiltonians for selected $\lambda$ values are diagonalized in
bases of decreasing size, as measured by ``$N_{\rm cut}$'',  which is
the cut-off  applied to the potential to study its decoupling
properties. (That is, the potential is set to zero for matrix
elements for which one or both states has $N > N_{\rm cut}$.) 
The left panel shows the results when only
the two-body evolved potential is used. The degree of decoupling
is measured by the point of departure from the asymptotic energy
for $\Nmax = 40$ as $N_{\rm cut}$ is lowered. As the potential is
evolved from the initial potential ($\lambda = \infty$) down to
$\lambda = 2.5$, decoupling is achieved for smaller spaces, which
means convergence is reached for smaller basis sizes. This is the
same pattern as found for realistic NN 
potentials~\cite{Jurgenson:2007td}. 

In the left panel of Fig.~\ref{fig:srg_3_body_decoupling},  the
ground-state energies asymptote to different values because the
evolution is not completely unitary. One might imagine that this
would affect the decoupling, but we show this is not the case
here. In the right panel, the induced three-body interaction is
kept, so the  curves asymptote to the same energy at large
$\Nmax$, while the same pattern of decoupling is observed. We
note that the  decoupling benefits afforded by evolution in the
oscillator basis are less straightforward than in the two-body
momentum basis studied for NN in \cite{Jurgenson:2007td}.   In
particular the cut-off errors  \emph{increase} for $\lambda$
smaller than the point at which the two-body-only binding energy
is at a minimum (i.e., for $\lambda = 1.5$ in
Fig.~\ref{fig:srg_3_body_decoupling}).


\subsection{Results for $A = 4$ and $A=5$}

Next we turn to $A=4$ and $A=5$, where we expect to see  the
effects of induced three-, four- and five-body forces. The key
issue is the relative sizes of these contributions;   we are
looking to test whether an initial hierarchy of few-body
interactions is preserved  and therefore can be truncated with a
controlled error.  (Of course, generalizing these results to
three dimensions will require repeating these tests for real 3D
nuclei.)

\begin{figure}
\begin{center}
\includegraphics*[width=6.4in]{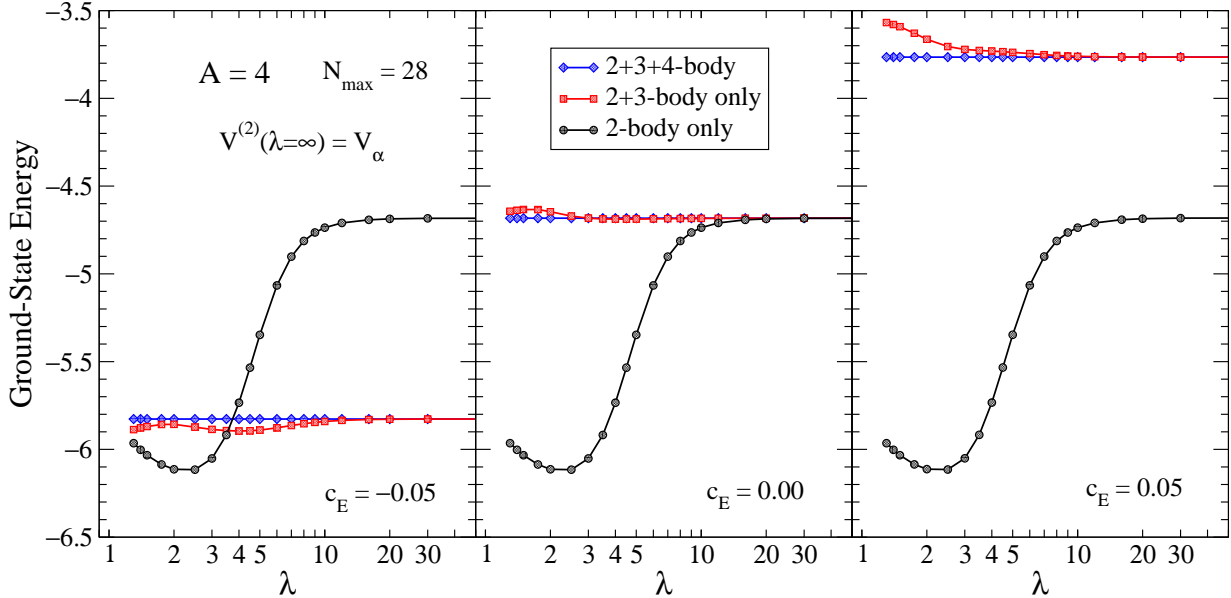}
\end{center}
\vspace*{-0.1in}
\caption{(color online) The lowest bound-state energy $E_4$ for a
four-particle system as a function of $\lambda$ with the inital
two-body potential $V_{\alpha}$ and different initial  three-body
force strengths ($c_E=\pm 0.05$).}
\label{fig:srg_4_body_Va}
\end{figure}

In applying the SRG in the four-particle space we have three
different calculations of the ground-state energy to compare. The
first is the two-body potential embedded successively in the
three- and four-particle spaces and then evolved  in the
four-particle space.  The resulting unitary transformations will
induce three- and four-body interactions that leave the
eigenvalues invariant. We can also evolve in the two-particle
space before embedding in the three- and four-particle spaces and
diagonalizing. As we saw before in Fig.~\ref{fig:srg_3_body} and
see now in Fig.~\ref{fig:srg_4_body_Va}, the two-body-only
evolution deviates because the Hamiltonian is not evolved by an
exactly unitary transformation. Finally, to find the relative size
of the three and four-body interactions we can evolve in the
three-particle space, thereby inducing only three-body forces.
Note that the two and three-body forces must be embedded
differently in the four particle space because they have
different combinatoric factors associated with them, i.e., there
are  ${4 \choose 2} = 6$ pairs and ${4 \choose 3} = 4$ triplets.
So, the proper mixture of two- and three-body force contributions
to the four-particle system interaction is $V = 6V^{(2)} +
4V^{(3)}$.

All three of these calculations for $A=4$ are shown in
Fig.~\ref{fig:srg_4_body_Va} for the two-body potential
$V_{\alpha}$ and several choices of the initial three-body
force.  The magnitude of $c_E$  was chosen so that the fractions
of the $A=3$ and $A=4$ ground-state energies from the three-body
interaction are roughly comparable to the corresponding fractions
for nuclei using typical realistic NN potentials. The qualitative
behavior is similar for other choices of $c_E$ and $V^{(2)}$.   
In all plots the curves for the two-body-only (black line with
circles), the two-plus-three (red line with squares), and the
full two-plus-three-plus-four interaction (blue line with
diamonds) show the hierarchy of different few-body interaction
components.  The difference between the square and diamond lines
represents the contribution of the four-body force, and the
difference between the circle and square lines is the contribution of the
three-body force alone. 

The four-body contribution is at most ten percent that of the
three-body, which is itself small compared to the two-body
contribution except when the latter gets small  for small
$\lambda$ (note the expanded scales on the figures).  Considering
calculations with different $c_E$ values, we see that the
$\lambda$ dependence of the induced four-body part depends on the
interplay of initial and induced forces. In some cases noticeable
(but small) evolution starts at $\lambda =10$ while in other
cases it is deferred until much smaller $\lambda$. Regardless of
the details, we stress that there is no sign that induced
many-body forces have rapid growth with $A$ or exhibit unusual
scaling.

We repeated for $A=4$ our test of decoupling that was shown in
Fig.~\ref{fig:srg_3_body_decoupling} for $A=3$. A similar pattern
of decoupling is found, namely an increased degree of decoupling
until  a $\lambda$ corresponding to the minimum of the
two-body-only ground-state energy of the $A=4$ system, after
which it deteriorates. 

\begin{figure}
\begin{center}
\includegraphics*[width=6in]{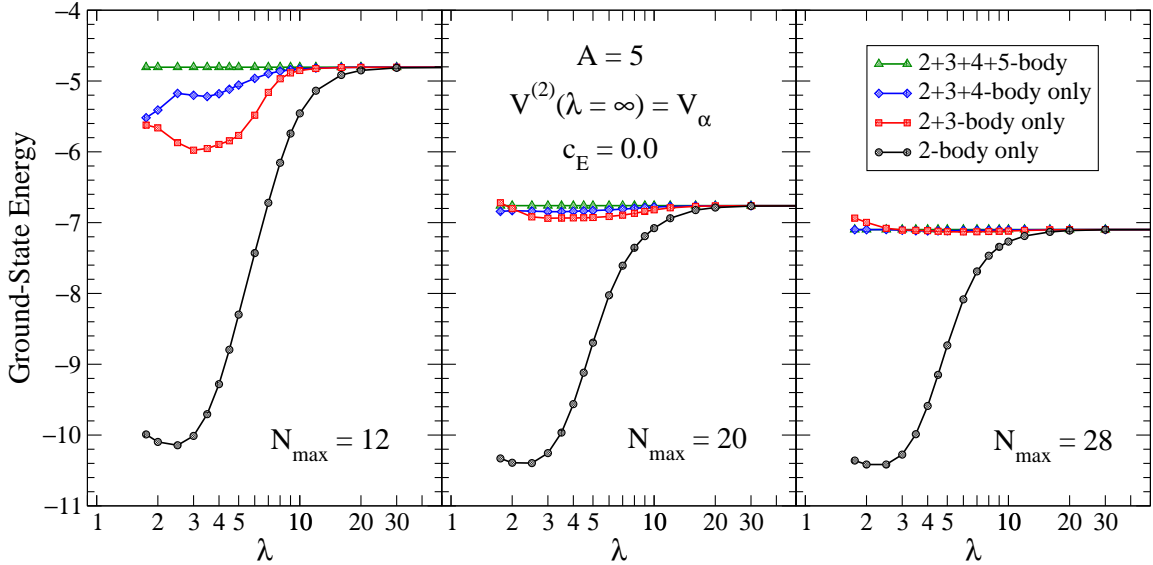}
\end{center}
\vspace*{-0.1in}
\caption{(color online) The lowest bound-state energy $E_5$ for a 5-particle system 
as a function of $\lambda$ with an initial two-body-only $V_{\alpha}$
potential for several values of $\Nmax$.}
\label{fig:srg_5_body}
\end{figure}

In Fig.~\ref{fig:srg_5_body} we show results for the SRG
evolution, with initial potential $V_{\alpha}$ and no initial
few-body interactions, in a five-particle system for several
values of $\Nmax$. The right panel, with $\Nmax = 28$, shows the
converged result. We see a decreasing hierarchy of induced
many-body contributions for all $\lambda$; the five-body
contribution is essentially negligible (or not distinguishable
from numerical noise). Differences at the lower $\Nmax$ sized
spaces arise both because the space needs to be large enough for
convergence to the exact energy eigenvalues but also because the
initial evolution of the potential needs a sufficiently large
space. Decoupling
may improve this feature but is dependent on the type of SRG
used and the basis in which it is implemented. This is important
to keep in mind for the generalization to the realistic nuclear
problem.


\subsection{Analysis of Three-Body Force Running}
\label{subsec:analysis}

\begin{figure}[ptb]
\includegraphics*[height=2.65in]{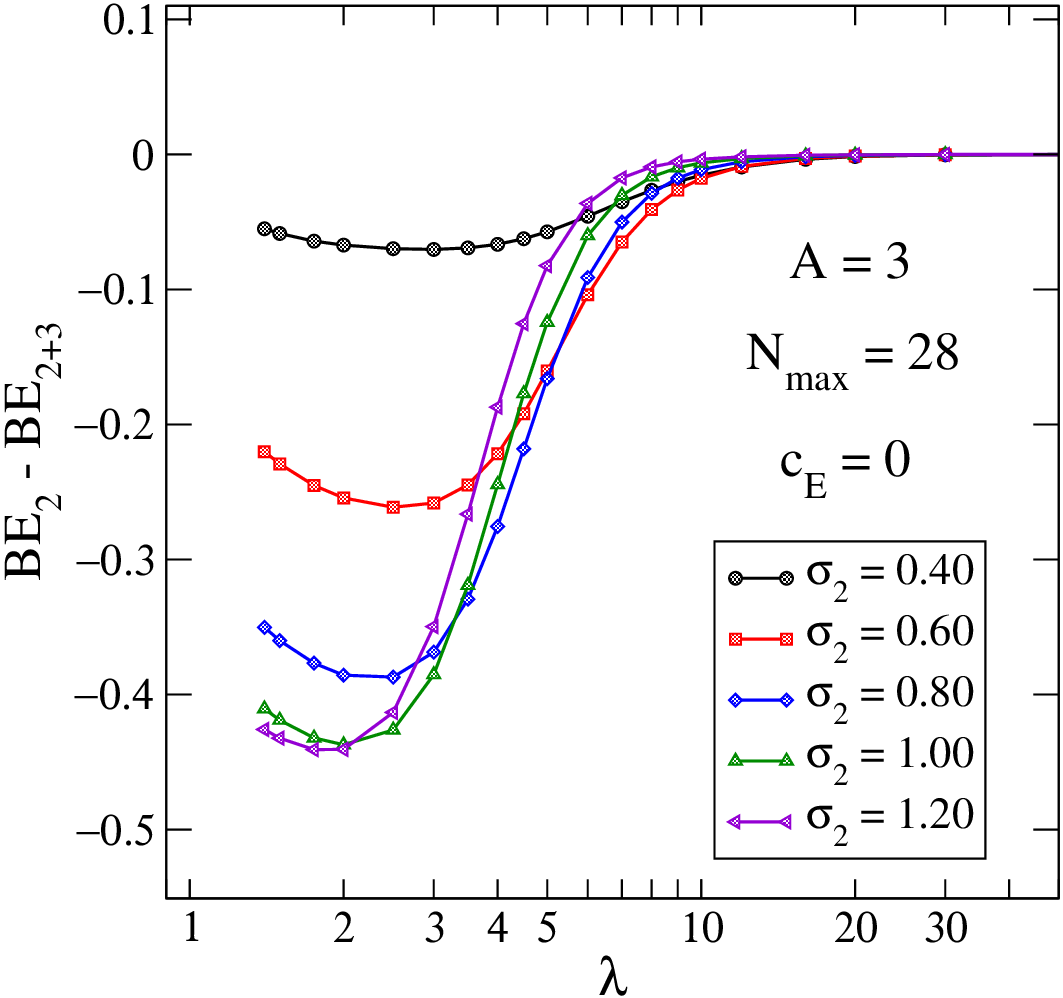}
  \hspace*{.1in}
\includegraphics*[height=2.65in]{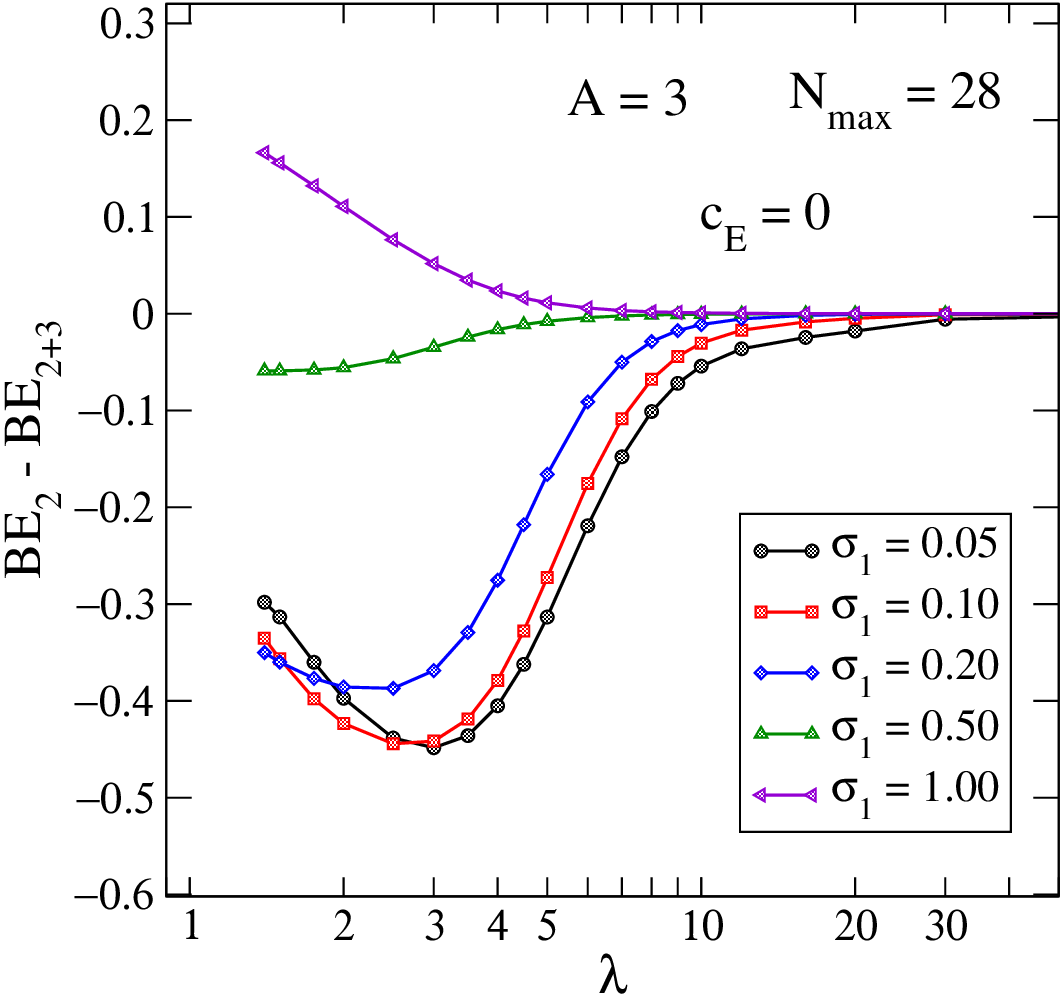}

\includegraphics*[height=2.65in]{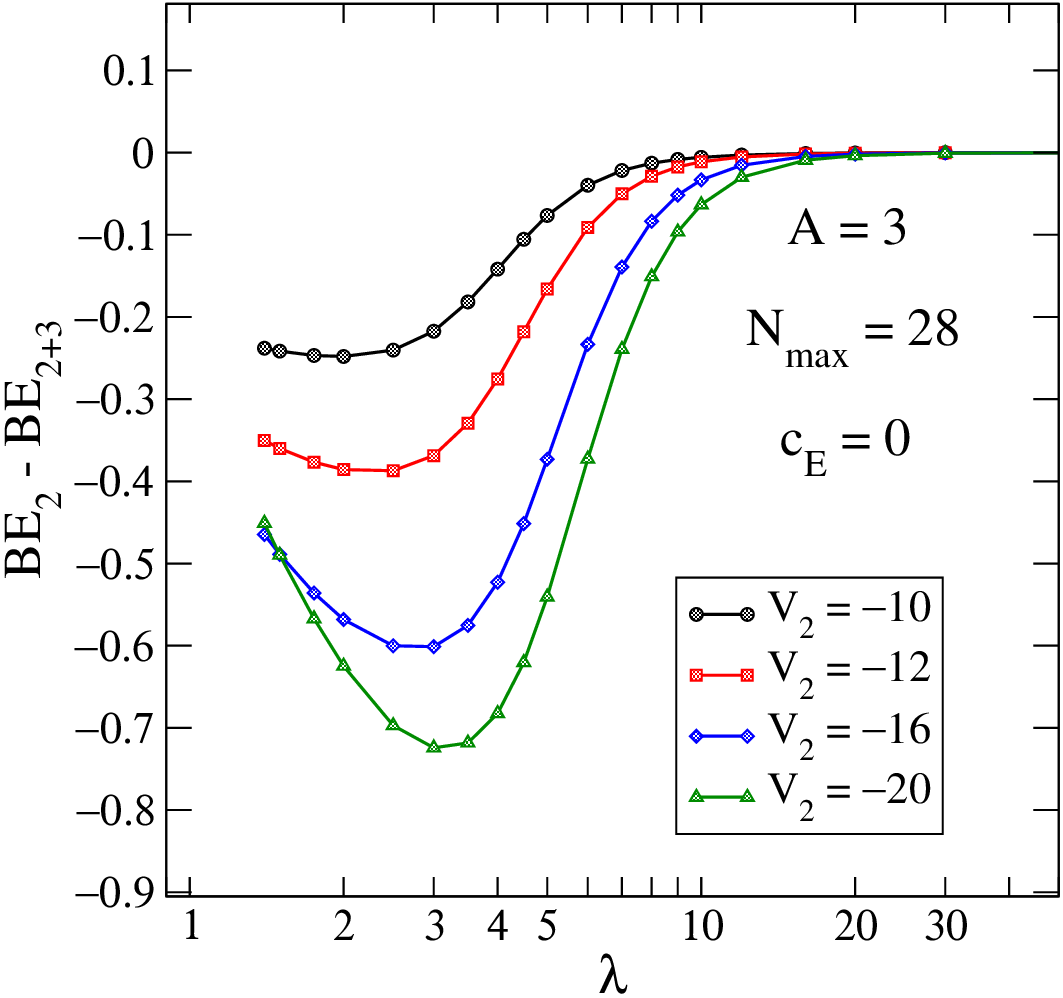}
  \hspace*{.1in}
\includegraphics*[height=2.65in]{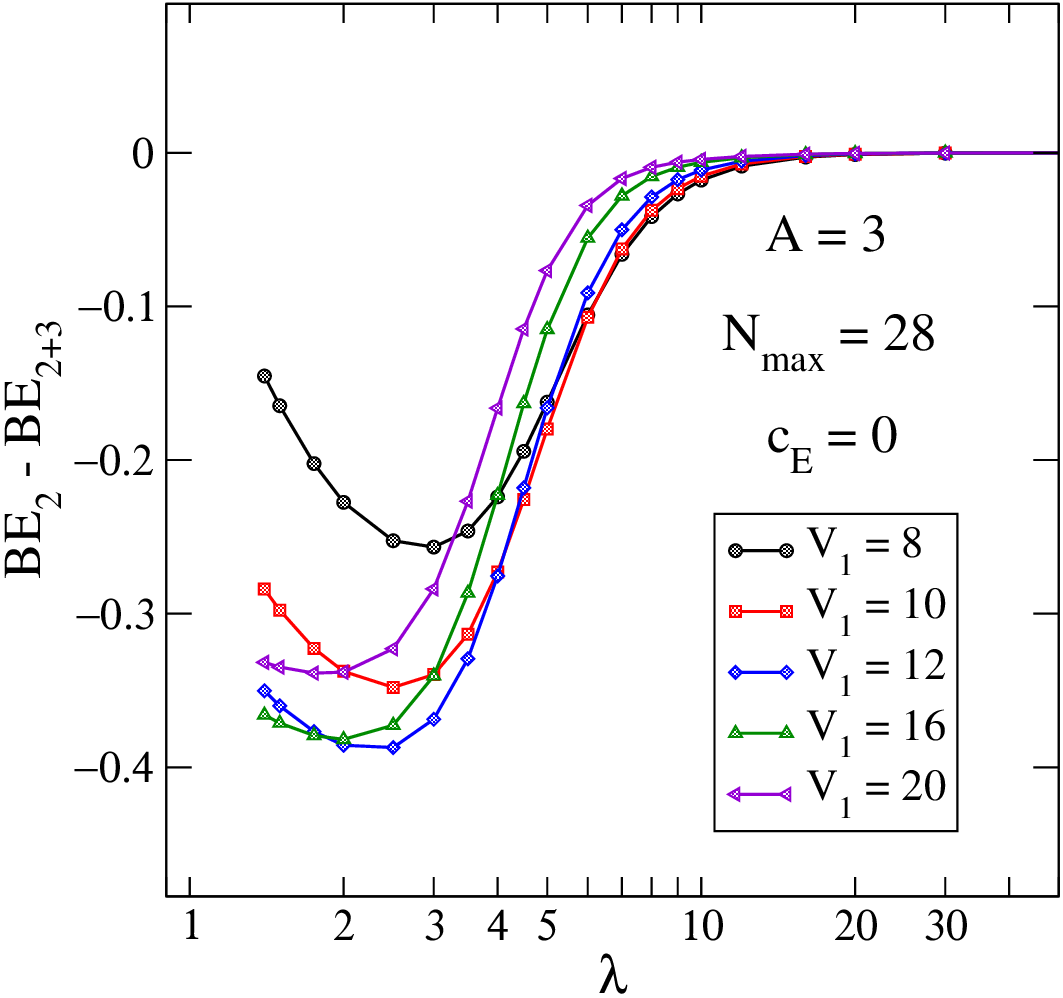}
\vspace*{-0.1in}
\caption{(color online) Differences of two-body-only and 
two-plus-three-body $A=3$ ground-state energies as a function of
$\lambda$. Each of the parameters of the potential $V_{\alpha}$
are varied in each plot as the other parameters are held
constant. The upper panels vary the ranges while the lower vary
the strengths; the left vary the attractive part and the right
vary the repulsive part.} 
\label{fig:3N_rel_err_Va}

\includegraphics*[height=2.65in]{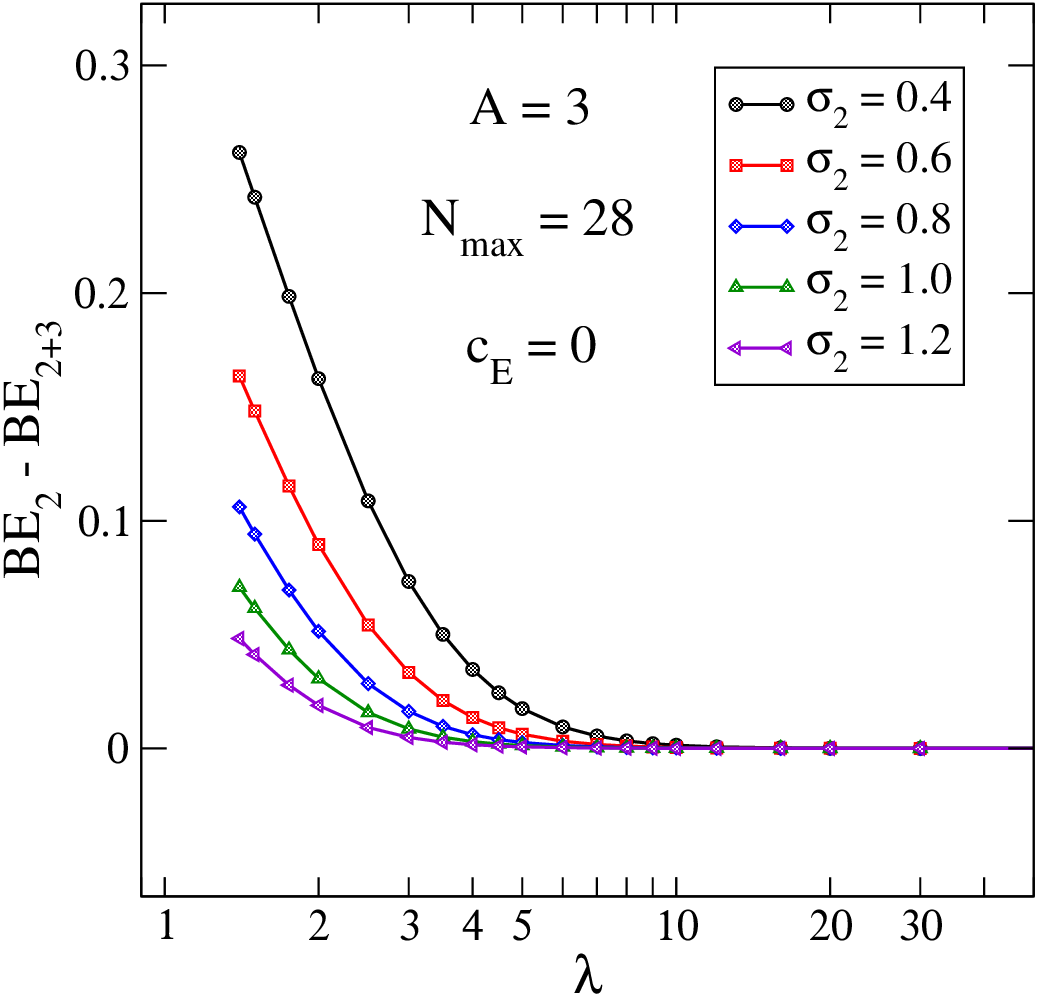}
  \hspace*{.1in}
\includegraphics*[height=2.65in]{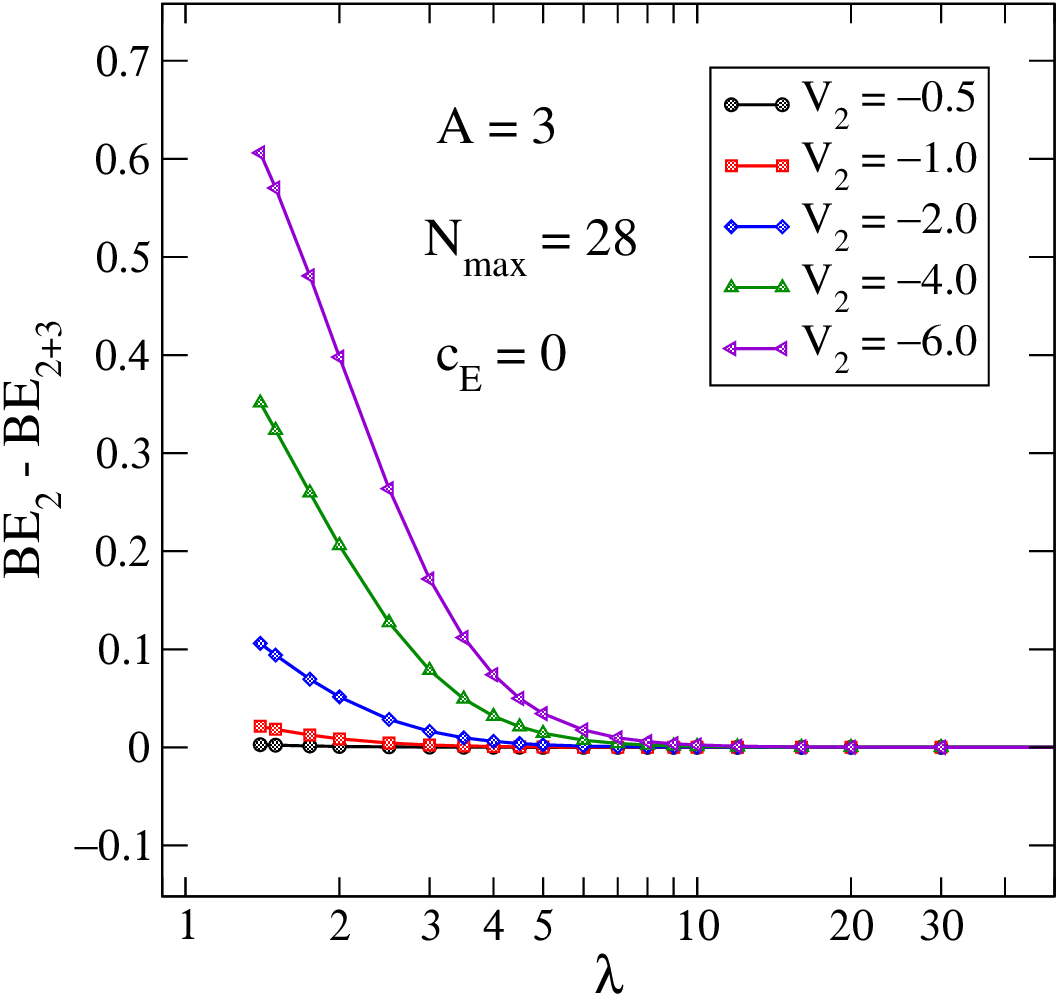}
\vspace*{-0.1in}
\caption{(color online) Same as Fig.~\ref{fig:3N_rel_err_Va} but for $V_{\beta}$.}
\label{fig:3N_rel_err_Vf}
\end{figure}

In this section we examine how  induced three-body interactions
evolve in our one-dimensional laboratory and make connections to
the diagrammatic expansion developed in
Ref.~\cite{Bogner:2007qb}. As raw material for this analysis,  we
plot in Fig.~\ref{fig:3N_rel_err_Va} the error of the evolving
two-body-only binding energy while varying several parameters of
the initial two-body interaction $V_{\alpha}$ for $A=3$
ground-state energies. In the left plots we vary the range (top)
and the strength (bottom) of the attractive part of
$V_{\alpha}$.  In the right plots we vary the range (top) and the
strength (bottom) of the repulsive part. In
Fig.~\ref{fig:3N_rel_err_Vf} we present similar plots for a
simpler system that starts with the initial attraction-only
potential $V_{\beta}$. Note that what is plotted shows the
evolution with $\lambda$  of the ground-state expectation value
of the three-body force.

Certain qualitative features are found as expected in these
figures. Shorter ranges imply enhanced
coupling from low to high momentum and therefore we anticipate
that the evolution will start sooner (i.e., at higher
$\lambda$).  This is seen clearly on the left in
Fig.~\ref{fig:3N_rel_err_Vf}  and for the variation of the
(shorter-ranged) repulsive potential in
Fig.~\ref{fig:3N_rel_err_Va} (top right plot). There is also an
unsurprising increase in the magnitude of the induced three-body
interaction at each $\lambda$ with increased magnitude of the potential, as
seen on the right in Fig.~\ref{fig:3N_rel_err_Vf} and on the
bottom left in Fig.~\ref{fig:3N_rel_err_Va}. For a more
definitive analysis we need to recall the discussion from
Ref.~\cite{Bogner:2007qb}.

\begin{figure}[bt]
\begin{center}
 \includegraphics*[width=5.5in]{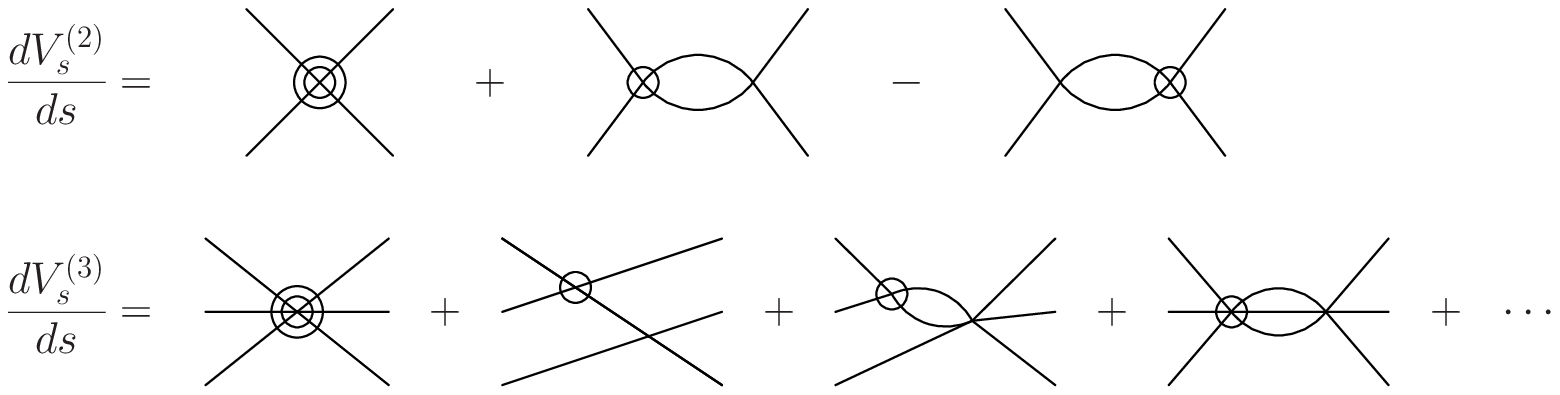}
\end{center}
\vspace*{-0.1in}
\caption{A diagrammatic decomposition of the SRG equation 
 (\ref{eq:commutator2}). A circle at a vertex denotes a
 commutator with $T_{\rm rel}$. }
\label{fig:srg_diagrams}
\end{figure}

The SRG evolution equation for the three-particle sector in
the notation of 
Ref.~\cite{Bogner:2007qb} is
\bea
  \frac{dV_{\flow}^{(2)}}{d\flow} +\frac{dV_{\flow}^{(3)}}{d\flow} 
  &=&  \mybar{\mybar{V}}_{\flow}^{(2)}+\mybar{\mybar{V}}_{\flow}^{(3)}+   
  [\mybar{V}_{\flow}^{(2)},V_{\flow}^{(2)}]
  +[\mybar{V}_{\flow}^{(2)},V_{\flow}^{(3)}]
  +[\mybar{V}_{\flow}^{(3)},V_{\flow}^{(2)}]
  +[\mybar{V}_{\flow}^{(3)},V_{\flow}^{(3)}]  
  \;,
  \label{eq:commutator2}
\eea 
where each bar denotes a commutator with $T_{\rm rel}$. We remind
the reader that $dT_{\rm rel}/ds = 0$ by construction. A
diagrammatic decomposition of this equation is shown in
Fig.~\ref{fig:srg_diagrams}. In the two-body sector, the
equation  reduces to the first term on the left and the first and
third terms on the right (the first row in 
Fig.~\ref{fig:srg_diagrams}). These terms keep two-particle
energy eigenvalues invariant under evolution. In the
three-particle sector, Eq.~(\ref{eq:commutator2}) results in not
only these two-body graphs with a disconnected spectator but
additional graphs involving connected combinations of two and
three-body interactions. The diagrams with two-body interactions
and a disconnected spectator line satisfy the two-body evolution
equations, and so will cancel out of the full three-particle-sector
evolution equation. Thus the evolution of the three-body
interaction is dictated by the connected diagrams (the second row
in Fig.~\ref{fig:srg_diagrams}). In summary, the evolution of the
$A$-body potential in the $A$-particle system is given by
\beqn 
  \frac{dV^{(A)}_s}{ds} = [\eta_s,H_s]_A \;, 
  \label{eq:srg_dVds}   
\eeqn
where the ``$A$'' subscript on the right side means the
fully connected $A$-particle terms.

To make a connection between the individual terms in the
three-body interaction evolution and the running of the ground-state
energy, we need to derive the evolution equations for the
\emph{expectation value} of $V^{(3)}_s$ in the ground state.
Denoting the ground-state wave function for the $A$-particle system
by $|\psi^A_s\ra$, it evolves according to 
\beqn
  |\psi^A_s\ra = U_s |\psi^A_{s=0}\ra \;,
  \quad \mbox \quad 
  \frac{d}{ds} |\psi^A_s\ra = \eta_s |\psi^A_s\ra
  \;,
\eeqn
where $U_s$ is the SRG unitary transformation at $s$ and
\beqn
  \eta_s = \frac{dU_s}{ds} U^\dagger_s = - \eta^\dagger_s
  \;.
\eeqn  
Then the matrix element of an operator $O_s$ evolves according to
\beqn
  \frac{d}{ds} \la \psi^A_s | O_s | \psi^A_s \ra
  = \la \psi^A_s |
  \frac{d O_s}{ds} - [\eta_s,O_s] 
    | \psi^A_s \ra
  \;.
  \label{eq:Oevolve}
\eeqn
If the operator $O_s$ evolves according to $O_s = U_s O_{s=0}
U_s^\dagger$, then the matrix element vanishes, as when $O_s = H_s$.

However, if we wish to see how one part of $H_s$
evolves, such as the expectation value of $V^{(3)}$, we obtain
\beqn
  \frac{d}{ds}\la \psi^A_s |V_s^{(3)}|\psi^A_s\ra = 
  \la \psi^A_s|\frac{dV^{(3)}_s}{ds} - [\eta_s,V_s^{(3)}]|\psi^A_s\ra
  \;,
\label{eq:dds_vev}
\eeqn
which does not give zero in general because $V_s^{(3)} \neq
U_sV_{s=0}^{(3)}U_s^{\dagger}$. 
In the two-particle case, the analog of
Eq.~\eqref{eq:dds_vev} gives $d\la V^{(2)} \ra/ds
= \la [\eta_s,T_{\rm rel}] \ra$. In the three-particle case, 
we can expand Eq.~\eqref{eq:dds_vev} as
\bea
\frac{d}{ds}\la \psi_s |V^{(3)}_s|\psi_s\ra &=& \la \psi_s|[\eta_s,H_s]_3 -
[\eta_s,V^{(3)}_s]|\psi_s\ra  \nonumber \\
&=& \la \psi_s |
   [\vbtr,T_{\rm rel}] + [\vbt,V^{(2)}_s]_c + [\vbt,V^{(3)}_s] + [\vbtr,V^{(2)}_s] +
[\vbtr,V^{(3)}_s] \nonumber \\
&& \qquad - [\vbt,V^{(3)}_s] - [\vbtr,V^{(3)}_s]  | \psi_s \ra  \nonumber \\
&=& \la \psi_s |  [\vbtr,H_s] + [\vbt,V^{(2)}_s]_c -
[\vbtr,V^{(3)}_s]| \psi_s \ra \nonumber \\
&=& \la \psi_s | [\vbt,V^{(2)}_s]_c - [\vbtr,V^{(3)}_s]| \psi_s \ra \;,
\label{eq:A3vevs}
\eea
where $\vbt$ and $\vbtr$ are the commutators $\vbt = [T_{\rm
rel},V^{(2)}_s]$ and $\vbtr = [T_{\rm rel},V^{(3)}_s]$. In the
third line, the expectation value of the commutator,
$[\vbtr,H_s]$, vanishes identically.

The term with the subscript ``$c$'' has had the two-body
disconnected diagrams removed. In our MATLAB implementation, this
subtraction is achieved by first embedding the two-particle-space
evolved version of this commutator in the three-particle space.
Computing $[\vbt,V^{(2)}]$ in the two-particle space alone
involves only the one-loop two-body interactions, so embedding 
in the three-particle sector results in only the disconnected
parts. This disconnected part can then be subtracted from the
total three-particle sector version of the same commutator,
leaving only the three-particle fully connected part.

\begin{figure}[bt!]
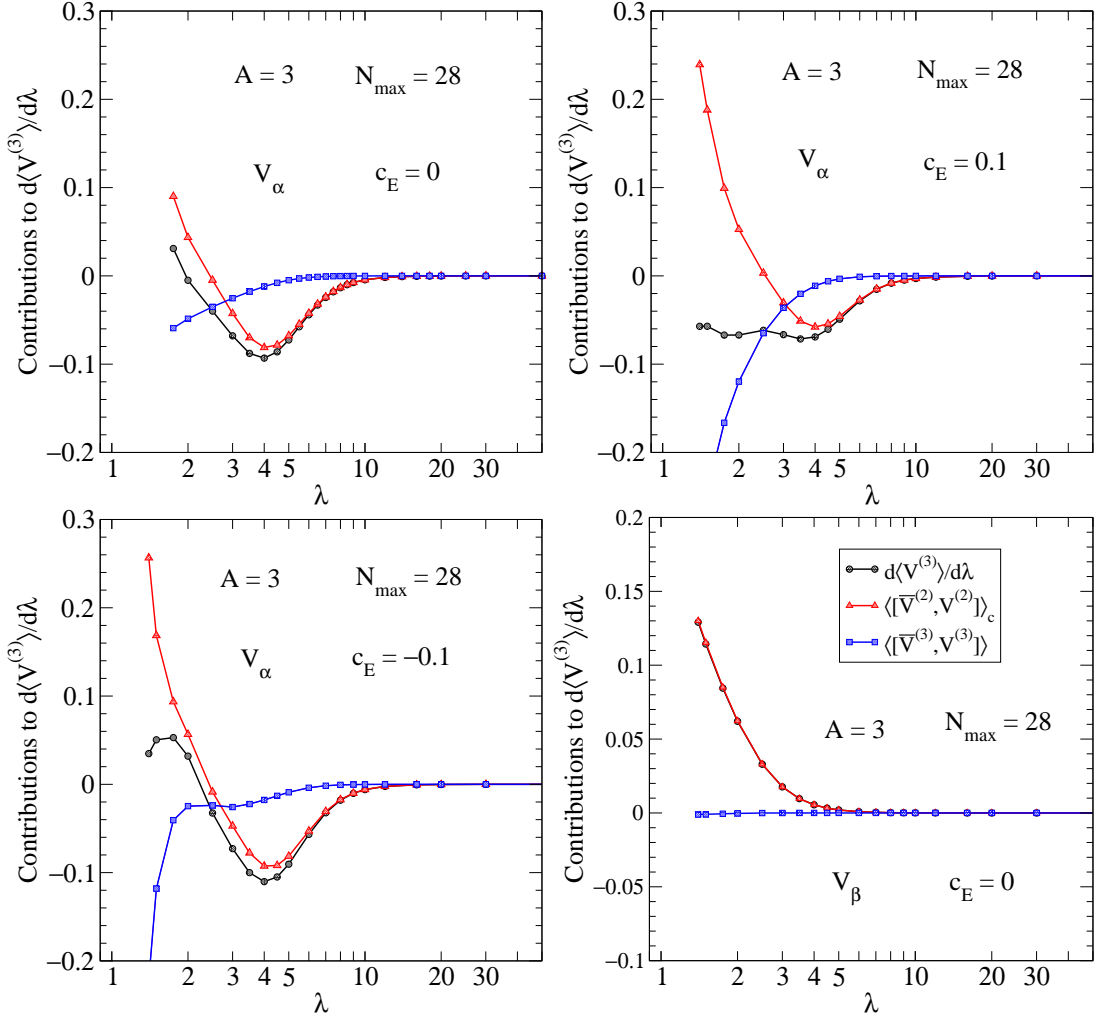

\begin{center}
\includegraphics*[height=2.65in]{VEV_Va_C0.eps}
\includegraphics*[height=2.65in]{VEV_Va_Cp1.eps}
\includegraphics*[height=2.65in]{VEV_Va_Cnp1.eps}
\includegraphics*[height=2.65in]{VEV_Vf_C0.eps}
\end{center}
\vspace*{-0.1in}
\caption{(color online) Contributions from individual terms to
the $A=3$ ground-state expectation value $d \la
V^{(3)}_\lambda\ra /d\lambda$ for several different two- and
three-body potentials, as indicated in the plots. We emphasize
that $\lambda \leq 2$ is very small, comparable to $\lambda \leq
1.5\,{\rm fm}^{-1}$ for NN forces in analogous calculations with the
NCSM~\cite{Bogner:2007rx}.}
\label{fig:srg_VEV}
\end{figure}

It is most useful for our analysis to convert from derivatives
with respect to $s$ to derivatives with respect to $\lambda$
using $\frac{d}{ds} = -\frac{\lambda^5}{4}\frac{d}{d\lambda}$. In
Fig.~\ref{fig:srg_VEV} we show the ground-state expectation
values of the right side of Eq.~\eqref{eq:dds_vev}, which are
broken down into the two terms from the right side of
Eq.~\eqref{eq:A3vevs} for $A=3$ and various potentials. It
is apparent that the drivers of three-body matrix element
evolution depend on the interplay between long- and short-range,
attractive and repulsive parts. The lower right panel of
Fig.~\ref{fig:srg_VEV} shows an increasing attractive strength of
the three-body force when starting from an attractive-only
two-body potential. In this regime the dominant contribution to
the evolution of the three-body potential matrix element is the tree-level
two-body connected part $[\vbt,V^{(2)}_s]_c$. This observation
accounts for the behavior in the right graph of
Fig.~\ref{fig:3N_rel_err_Vf}, where the size of the error scales
(roughly) like $(V^{(2)})^2$. Varying the long-range attraction
strength in Fig.~\ref{fig:3N_rel_err_Va} shows a similar effect.

More generally, the impact on ground-state energies of the
induced three-body interaction depends on the details of the
correlations in the wave function. The other plots in
Fig.~\ref{fig:srg_VEV} for the more realistic initial two-body
potential, $V_{\alpha}$, show the interplay between two- and
three-body contributions to the three-body matrix element
evolution. The three-body contribution to the three-body
evolution stays small until the longer-range attractive part of
the potential begins to affect the evolution. Most of the change
is from $\lambda = 8$ to $\lambda = 3$, which is dominated by
$\la \overline{V}^{(2)},V^{(2)}\ra$. Thus, the feedback of the
three-body potential depends on the initial conditions, but
in general insures that the binding energy contribution stays
small.

We can repeat the above analysis for $A \geq 4$ and find no fully
connected terms with only two-body forces. Again, disconnected
terms involving two and three body potentials cancel out in the
lower sectors. The leading terms are commutators with one
$V^{(2)}_s$ and one $V^{(3)}_s$, followed by connected terms
quadratic in $V^{(3)}_s$ and one term quadratic in $V^{(4)}_s$.
All terms are small and additional cancellations among them
further suppress the four-body contribution. Thus, the initial
hierarchy of many-body forces implies that induced four-body (and
higher-body) forces will be small.


\section{Summary}
  \label{sec:conclusion}

In this paper we use a one-dimensional system of bosons with
short-range repulsion and mid-range attraction as a laboratory to
explore the evolution of many-body forces with the SRG. These
calculations serve as a proof-of-principle that induced few-body
interactions can be calculated without solving T-matrix
equations, as necessary in other renormalization group
approaches. They establish that working within a harmonic
oscillator basis is practical, although the generalization to
three dimensions will be much more computationally intensive. 
This generalization is direct, however, because the recursive
construction of properly symmetrized few-body basis states used
here has already been carried out in the context of no-core shell
model  calculations~\cite{NCSM1a,NCSM1b,NCSM1c}. The first steps
with an $A=3$ nuclear system will be taken soon, with the
resulting three-body matrix elements then used as input to
calculations in larger nuclei.

The patterns of SRG evolution observed in our one-dimensional
laboratory have many similaries to those observed in three
dimensions with realistic nuclear interactions.  This includes,
for example, decoupling in few-body systems, which was studied
for nuclei with two-body forces only~\cite{Jurgenson:2007td}. Here
we find decoupling is unchanged when induced three-body
interactions are included.   Results for the contributions of
induced three-body forces to  ground-state energies show the same
pattern of running observed for nuclear forces when a model
two-body  potential with short-range repulsion and long-range
attraction is used. Differences in the running when an
attractive-only two-body potential is used are understood in
terms of the dominant contribution to the evolution equation for
the three-body potential. Starting from an initial decreasing
hierarchy of forces (two-body larger than three-body and all
higher-body forces zero), the hierarchy is maintained for useful
ranges of the SRG evolution. Induced four- and five-body forces
are found to be very small. This is encouraging for the
applications to realistic nuclei, but of course is not
conclusive.

In a future study,  we will use the one-dimensional laboratory to test
alternative methods to evolve many-body interactions. These
include the use of a Slater determinant basis of harmonic
oscillators (``$m$ scheme'') and evolution in momentum space
applying the diagrammatic methods from Ref.~\cite{Bogner:2007qb}.
With the oscillator basis methods we will also consider evolving
with a choice of the SRG generator  $G_s$ that is diagonal in the
basis (i.e., $G_s=H_{osc}$). The block diagonalization
characteristic of ``$\vlowk$'' potentials (as opposed to band
diagonalization as studied here) has been demonstrated for $A=2$
in the SRG using a different choice for the flow operator $G_s$. 
We will test whether this can be extended to $A \geq 3$, which
would circumvent the need to solve coupled T-matrix problems.  We
will also test an alternative to explicit few-body forces, which
is to use a normal-ordering prescription in the SRG to generate
``density-dependent'' two-body interactions.  Finally, another
avenue to explore is the evolution of other operators in few-body
systems.

\medskip

This work was supported in part by the National Science 
Foundation under Grant No.\ PHY--0653312 and the UNEDF SciDAC
Collaboration under DOE Grant DE-FC02-07ER41457. We thank
E. Anderson, S. Bogner, R. Perry, L. Platter, J. Drut, and A. Schwenk
for their comments and discussions.

\appendix
\section{Transformation Brackets}
\label{app:trans_brack}

Transformation brackets are the expansion coefficients in the
oscillator basis of one  system of coordinates in terms of
another~\cite{moshinsky,shlomo}. We apply them to relate two different choices of Jacobi
coordinates. Here, we show the relevant transformation using the
three-particle harmonic  oscillator states defined in
Eq.~\eqref{eq:ho_basis} and then generalize at the end.

The single particle momenta are $k_1$, $k_2$, and $k_3$. The
unprimed Jacobi momenta [see Eq.(~\ref{eq:Jacobi_coords})] are 
\bea
p_1 &=&  \frac{1}{\sqrt{2}} (k_1 - k_2) \;, \nonumber \\
p_2 &=& \sqrt{\frac{2}{3}} ((k_1+k_2)/2 - k_3) \;,
\eea
and the primed coordinates are obtained from exchanging $k_2$ and
$k_3$:
\bea
p_1' &=&  \frac{1}{\sqrt{2}} (k_1 - k_3) \;, \nonumber \\
p_2' &=& \sqrt{\frac{2}{3}} ((k_1+k_3)/2 - k_2) \;.
\eea
After some algebra, the transformation that exchanges the last
two particles (i.e., $k_2$ and $k_3$) can be written as
\bea
  \left( \begin{array}{c} p_1' \\ p_2' \end{array}\right) 
   &=& \left( \begin{array}{cc} \frac{1}{2} & \frac{\sqrt{3}}{2} 
   \\[4pt]
  \frac{\sqrt{3}}{2} & -\frac{1}{2}  \end{array}\right) 
  \left( \begin{array}{c} p_1 \\ p_2 \end{array}\right) \;.
  \label{eq:p_transformation}
\eea
which enables us to express the primed oscillator states in terms
of the unprimed ones. 

We denote the three-particle oscillator basis by $|n_1n_2\ra =
\eta_1^{\dagger}\eta_2^{\dagger}|0\ra$ where we have set
$\hbar\omega = 1$ for simplicity in this appendix. The
transformation that exchanges the last two single-particle
coordinates can again be written as
\bea
  \left( \begin{array}{c} \eta_1' \\ \eta_2' \end{array}\right) 
   &=& \left( \begin{array}{cc} \frac{1}{2} & \frac{\sqrt{3}}{2}  
   \\[4pt] 
  \frac{\sqrt{3}}{2} & -\frac{1}{2}  \end{array}\right) 
  \left( \begin{array}{c} \eta_1 \\ \eta_2 \end{array}\right) \;,
  \label{eq:eta_transformation}
\eea
The derivation of the
harmonic oscillator transformation bracket follows directly as
\newcommand{\Anorm}{\frac{1}{\sqrt{n_1!n_2!n_1'!n_2'!}}}
\bea
  \la n_1'n_2' | n_1n_2 \ra_3 &=& \la 0| \Anorm
   \eta_1'^{n_1'}\eta_2'^{n_2'} \eta_1^{\dagger n_1} 
   \eta_2^{\dagger n_2}|0\ra \nonumber \\ 
&=& \la 0|\Anorm \Bigl[\frac{\eta_1}{2} 
   + \frac{\sqrt{3}\eta_2}{2}\Bigr]^{n_1'} \nonumber \\
&& \hspace{2cm}  \null\times \Bigl[\frac{\sqrt{3}}{2}\eta_1 -
  \frac{1}{2}\eta_2\Bigr]^{n_2'} 
  \eta_1^{\dagger n_1} \eta_2^{\dagger n_2}  |0\ra \nonumber \\
&=& \la 0|\Anorm  \sum_{k=0}^{n_1'} {n_1' \choose k}
  \Bigl[\frac{1}{2}\eta_1\Bigr]^{n_1'-k}
  \Bigl[\frac{\sqrt{3}}{2}\eta_2\Bigr]^k 
     \nonumber \\
&& \hspace{2cm}  \null\times\sum_{j=0}^{n_2'} {n_2' \choose j}
 \Bigl[\frac{\sqrt{3}}{2}\eta_1\Bigr]^{n_2'-j}
 \Bigl[-\frac{1}{2}\eta_2\Bigr]^j 
\eta_1^{\dagger n_1}\eta_2^{\dagger n_2} |0\ra \nonumber \\
&=& \Anorm \sum_{k=0}^{n_1'} \sum_{j=0}^{n_2'} 
{n_1' \choose k}{n_2' \choose j} 
\biggl[\frac{1}{2}\biggr]^{n_1'-k+j}
\biggl[\frac{\sqrt{3}}{2}\biggr]^{n_2'-j+k}(-1)^j \nonumber \\
&& \hspace{2cm}  \null\times n_1!n_2!\delta_{n_1'-k+n_2'-j,n_1}
\delta_{k+j,n_2} \nonumber \\
&=& \sqrt{\frac{n_1!n_2!}{n_1'!n_2'!}} 
\sum_{k=0}^{n_1} {n_1' \choose k}{n_2' \choose n_2-k}
\biggl[\frac{1}{2}\biggr]^{n_1'+n_2-2k}
\biggl[\frac{\sqrt{3}}{2}\biggr]^{n_2'-n_2+2k}(-1)^{n_2-k} \;.
\label{eq:trans_bracket}
\eea
The second line is obtained from operating the transformation
on the creation operators  $\eta_s^\dagger$. The third line is
the application of the binomial theorem. The fourth  balances the
oscillator creation and annihilations, and the fifth is just some
simplification. 

In the general A-particle system the transformation to exchange
the last two particles, $k_{A-1}$ and $k_A$, can be written as
\bea
\left( \begin{array}{c} \eta_{A-2}' \\ \eta_{A-1}' \end{array}\right) 
 &=& \left( \begin{array}{cc} 
  \sqrt{\frac{1}{d+1}} & \sqrt{\frac{d}{d+1}}  \\[4pt] 
  \sqrt{\frac{d}{d+1}} & -\sqrt{\frac{1}{d+1}}  \end{array}\right) 
\left( \begin{array}{c} \eta_{A-2} \\ \eta_{A-1} \end{array}\right) \;,
\label{eq:etap_trans}
\eea
where $d = (A-1)^2-1$ is the number of generators of the rotation
group, $U(A-1)$, or the group $U(A)$ with the center of mass
coordinate held fixed.  An expression for the bracket $\la
n_{A-2}n_{A-1} | n_{A-2}'n_{A-1}' \ra_{A(A-2)}$, which appears in
Eq.~(\ref{eq:symmetrizer_A}), is obtained from
Eq.~(\ref{eq:trans_bracket}) by substituting the general
coordinate transformation Eq.~(\ref{eq:etap_trans}) for the
three-particle transformation Eq.~(\ref{eq:eta_transformation}),
or $\sqrt{1/(d+1)}$ and $\sqrt{d/(1+d)}$ for $1/2$ and
$\sqrt{3}/2$.


\end{document}